\documentclass[aps, prd, letterpaper, amsmath, amssymb, superscriptaddress, twocolumn, floatfix, nofootinbib]{revtex4-1}
\usepackage[pdftex]{graphicx}
\usepackage[pdftex, pdfstartview={FitH}, pdfnewwindow=true, colorlinks=true, citecolor=blue, filecolor=blue, linkcolor=blue, urlcolor=blue, pdfpagemode=UseNone, bookmarks=false]{hyperref}

\def\muhat{\hat{\mu}}
\def\nuhat{\hat{\nu}}

\def\rhohat{\hat{\rho}}
\newcommand{\vev}[1]{\ensuremath{\left\langle #1 \right\rangle} }
\newcommand{\nn}{\nonumber}

\begin{document}
\title{Novel phases in strongly coupled four-fermion theories}

\author{Simon Catterall}
\email{smcatter@syr.edu}
\affiliation{Department of Physics, Syracuse University, Syracuse, New York 13244, United States}
\author{David Schaich}
\affiliation{Department of Physics, Syracuse University, Syracuse, New York 13244, United States}
\affiliation{AEC Institute for Theoretical Physics, University of Bern, 3012 Bern, Switzerland}

\date{1 July 2017}

\begin{abstract}
We study a lattice model comprising four massless reduced staggered fermions in four dimensions coupled through an $SU(4)$-invariant four-fermion interaction.
We present both theoretical arguments and numerical evidence that no bilinear fermion condensates are present for any value of the four-fermi coupling, in contrast to earlier studies of Higgs--Yukawa models with different exact lattice symmetries.
At strong coupling we observe the formation of a four-fermion condensate and a mass gap in spite of the absence of bilinear condensates.
Unlike those previously studied systems we do not find a ferromagnetic phase separating this strong-coupling phase from the massless weak-coupling phase.
Instead we observe long-range correlations in a narrow region of the coupling, still with vanishing bilinear condensates.
While our numerical results come from relatively small lattice volumes that call for caution in drawing conclusions, if this novel phase structure is verified by future investigations employing larger volumes 
it may offer the possibility for new continuum limits for strongly interacting fermions in four dimensions.
\end{abstract}

\maketitle

\section{Introduction}
In this paper we study a four-dimensional lattice theory
comprising four massless reduced staggered fermions coupled through an $SU(4)$-invariant four-fermion interaction. Strong-coupling arguments indicate
that the system develops a massive phase for sufficiently large four-fermi coupling without breaking symmetries.  Such a (paramagnetic strong-coupling or PMS) phase has been
seen before in other lattice Higgs--Yukawa models, and is generically separated from a massless
paramagnetic weak-coupling (PMW) phase by an intermediate ferromagnetic
phase characterized by a symmetry-breaking bilinear fermion condensate.
A representative small sample of this earlier work can be found in Refs.~\cite{Stephenson:1988td, Hasenfratz:1988vc, Lee:1989xq, Lee:1989mi, Bock:1990cx, Hasenfratz:1991it, Golterman:1992yha} and references therein.
The key result of our current work is that we see no evidence for this intermediate broken phase in the model described here, which possesses different exact lattice symmetries than the systems considered previously.  Instead we observe a narrow region of four-fermi coupling separating the
PMW and PMS phases in which the fermions develop long-range
correlations but no bilinear condensate is formed.

The same model was studied previously in three dimensions utilizing
three different numerical algorithms: fermion bags, rational hybrid Monte Carlo (RHMC) and quantum
Monte Carlo~\cite{Ayyar:2014eua, Ayyar:2015lrd, Catterall:2015zua, He:2016sbs}.
These studies revealed an interesting two-phase structure for the model; a massless
phase at weak coupling (the analog of the PMW phase in four dimensions) is separated by a continuous
phase transition with non-Heisenberg exponents from a massive (PMS-like) phase at strong coupling.

The four-dimensional theory which is the focus of the current work
was also recently studied in Ref.~\cite{Ayyar:2016lxq}. The conclusion of that work
was that a narrow broken phase reappears upon lifting the
theory from three to four dimensions. It is important to note that this conclusion was not based on
an explicit measurement of a condensate but rather inferred from the volume scaling of a certain susceptibility.

In our work we have augmented the action used in that recent study with source terms to directly address the question
of whether spontaneous symmetry breaking associated with the formation of
specific bilinear condensates takes place.  Our measurements of susceptibilities at zero source agree with those
reported in \cite{Ayyar:2016lxq} and are consistent with the possibility of a narrow intermediate phase that they describe.
However, in contrast to that work we do not see any evidence for the formation of bilinear condensates and spontaneous
symmetry breaking in that region of the phase diagram.  Thus the intermediate phase---if it exists---must be of
an unusual character.  It goes without saying that the appearance of potentially new critical behavior in
lattice theories of strongly interacting fermions in four dimensions would be very interesting from both
theoretical and phenomenological viewpoints, with regards to constructing theories of BSM physics.
Similar models have also received considerable interest in recent years within the condensed matter community~\cite{Fidkowski:2009dba, Morimoto:2015lua}.
Of course one must be somewhat cautious in drawing too strong a conclusion at this point since our simulations are currently limited to rather modest lattice volumes.
We plan to investigate larger volumes in future RHMC calculations, and also hope to see additional studies of this system employing fermion bags or other algorithms.

The plan of the paper is as follows: in the next section we describe the lattice model and its
symmetries and in Sec.~\ref{sec:strong} we describe
the phases expected at strong and weak four-fermi
coupling. In Sec.~\ref{sec:aux} we show how to replace the four-fermion interaction by appropriate Yukawa
terms and prove that the resulting Pfaffian is real positive semi-definite. This fact allows
us to simulate the model using the RHMC algorithm and we show results for the phase diagram from
those simulations in Sec.~\ref{sec:phases}. To examine the question of whether spontaneous symmetry breaking occurs we have conducted the bulk of our simulations with an action that includes explicit symmetry-breaking source terms
and we include a detailed study of the volume and source dependence of possible bilinear condensates in Sec.~\ref{sec:cond}.
In Sec.~\ref{sec:CW} we strengthen these conclusions by computing the one-loop Coleman--Weinberg effective potential associated with a particular single-site condensate that breaks the $SU(4)$ symmetry of the model.
We show that the unbroken state remains a minimum of the potential for all values of the four-fermi coupling, in agreement with our numerical study.
Finally we summarize our findings and outline future work in Sec.~\ref{sec:concl}.

\section{Lattice action and symmetries}

Consider a theory of four {\it reduced} staggered fermions in four dimensions
whose action contains a single-site $SU(4)$-invariant
four-fermion term.\footnote{The $SO(4)$ symmetry discussed in \cite{Catterall:2015zua} naturally
enhances to $SU(4)$ if the fermions are allowed to be complex. Such an enlargement of the
symmetry group does not invalidate the arguments needed to construct an auxiliary field representation
or to show the Pfaffian is real and positive semi-definite.}
The action is
\begin{equation}
  \begin{split}
    \label{action}
    S = & \sum_x\sum_\mu \eta_\mu(x) \psi^a(x)\Delta^{ab}_\mu\psi^b(x) \\
        & - \frac{1}{4}G^2 \sum_x \epsilon_{abcd}\psi^a(x)\psi^b(x)\psi^c(x)\psi^d(x)
  \end{split}
\end{equation}
where $\Delta^{ab}_\mu\psi^b(x)=\frac{1}{2}\delta_{ab}\left(\psi^b(x+\muhat)-\psi^b(x-\muhat)\right)$ with
$\muhat$ representing unit displacement in the lattice in the $\mu$ direction
and $\eta_\mu(x)$ is the usual staggered fermion phase $\eta_\mu(x)=\left(-1\right)^{\sum_{i=0}^{\mu-1} x_i}$.
The reduced staggered
fermions are taken to transform according to
\begin{equation}
\psi(x) \to e^{i\epsilon(x) \alpha}\psi(x)\end{equation}
with $\alpha$ an arbitrary element of the algebra of $SU(4)$ and
$\epsilon(x)=\left(-1\right)^{\sum_{i=0}^{d-1}x_i}$ denoting the lattice parity. The presence of the four-fermion interaction breaks the usual global $U(1)$ symmetry
down to $Z_4$ whose action is given explicitly by $\psi\to \Gamma \psi$ where $\Gamma=\left[1,-1,i\epsilon(x),-i\epsilon(x)\right]$.
The action is also invariant under the shift symmetry
\begin{equation}
\psi(x)\to \xi_\rho(x)\psi(x+\rhohat)\end{equation}
where the flavor phase $\xi_\mu(x)=\left(-1\right)^{\sum_{i=\mu+1}^{d-1}x_i}$.
These shift symmetries can be thought of as a discrete
remnant of continuum chiral symmetry~\cite{Bock:1992yr}.

These symmetries strongly constrain the possible bilinear terms that can arise in the lattice effective action
as a result of quantum corrections. For example,
a single-site mass term of the form $\psi^a(x)\psi^b(x)$
breaks the $SU(4)$ invariance and the $Z_4$ symmetry
but maintains the shift symmetry, while $SU(4)$-invariant
bilinear terms constructed from products of
staggered fields within the unit hypercube generically break the
shift symmetries~\cite{vandenDoel:1983mf, Golterman:1984cy}.\footnote{The usual single-site mass term $\overline{\psi}^{a}(x)\psi^a(x)$ that is possible for a full staggered field
is invariant under all symmetries but this
term is absent for a {\it reduced} staggered field since in this case there is no independent $\overline{\psi}$ field.}
The possible $SU(4)$-invariant multilink bilinear operators for a reduced staggered fermion are
\begin{align}
  O_1 & = \sum_{x,\mu} m_\mu\epsilon(x)\xi_\mu(x)\psi^a(x)S_\mu\psi^a(x)                                  \label{ops} \\
  O_3 & = \sum_{x,\mu,\nu,\lambda} m_{\mu\nu\lambda} \xi_{\mu\nu\lambda}(x) \psi^a(x)S_\mu S_\nu S_\lambda \psi^a(x) \nn
\end{align}
where $\xi_{\mu\nu\lambda}(x) \equiv \xi_\mu(x)\xi_\nu(x+\muhat)\xi_\lambda(x+\muhat+\nuhat)$ and $m_{\mu\nu\lambda}$ is totally antisymmetric in its indices. In these expressions the
symmetric translation operator $S_\mu$ acts on a lattice field according to $S_\mu \psi(x)=\psi(x+\muhat)+\psi(x-\muhat)$.

Notice that while the exact lattice symmetries constrain the form of the
effective action of the theory it is still possible for condensates of either the single-site and/or multilink operators to appear if the vacuum state spontaneously breaks
one or more of these symmetries.

\section{\label{sec:strong}Strong-coupling behavior}

Before turning to the auxiliary field
representation of the four-fermi term and our numerical simulations we
can first attempt to understand the behavior of the theory in the limits of both weak and strong coupling.
At weak coupling one expects that the fermions are massless and there should be
no bilinear condensate since the four-fermi term is
an irrelevant operator by power counting.

In contrast the
behavior of the system for large coupling can be deduced from a strong-coupling expansion.
The leading term corresponds to the static limit $G\to\infty$ in which the kinetic operator is dropped
and the exponential of the four-fermi term is expanded in powers of $G$.
In this limit the partition function for lattice volume $V$ is saturated by terms of the form
\begin{equation}
  \label{eq:strong}
  \begin{split}
    Z & \sim \Bigg[6G^2 \int d\psi^1(x)d\psi^2(x) d\psi^3(x) d\psi^4(x) \\[-10 pt]
      & \hspace{75 pt} \times \psi^1(x)\psi^2(x)\psi^3(x)\psi^4(x)\Bigg]^V
  \end{split}
\end{equation}
corresponding to a single-site four-fermi condensate. To leading order in this
expansion it should also be clear that
the vev of any bilinear operator will be zero since one cannot then saturate all the Grassmann integrals using
just the four-fermion operator.

To compute the fermion propagator at strong coupling it is convenient to rescale the
fermion fields by $\sqrt{\alpha}$ where $\alpha=\frac{1}{\sqrt{6}G} \ll 1$ which removes the coupling from the interaction term and instead
places a factor of $\alpha$ in front of the
kinetic term.  To leading order in $\alpha$ the partition function is now unity. The strong-coupling
expansion then corresponds to an expansion in $\alpha$. We follow the procedure described
in~\cite{Eichten:1985ft} and
consider the fermion propagator $F(x)=\vev{\psi^1(x)\psi^1(0)}$. To integrate out the fields
at site $x$ one needs to bring down
$\psi^2(x)$, $\psi^3(x)$, $\psi^4(x)$ from the kinetic term. This yields a leading contribution
\begin{widetext}
\begin{equation}
  F(x) = \left(\frac{\alpha}{2}\right)^3 \int_x D\psi \sum_\mu\eta_\mu(x) \left(\Psi^1(x+\muhat)-\Psi^1(x-\muhat)\right)\psi^1(0)e^{-S}
\end{equation}
where $\Psi^1=\psi^2\psi^3\psi^4$ and $\int_x$ means we no longer include an integration
over the fields at $x$. We then repeat this procedure at $x\pm \muhat$ leading to
\begin{equation}
  F(x) = \left(\frac{\alpha}{2}\right)^3\sum_\mu\eta_\mu(x)\left(\delta_{x+\muhat,0}-\delta_{x-\muhat,0}\right) + \left(\frac{\alpha}{2}\right)^4\int_{x,x\pm \muhat} D\psi \sum_\mu\left( \psi^1(x+2\muhat)+\psi^1(x-2\muhat)\right)\psi^1(0)e^{-S}.
\end{equation}
\end{widetext}
Notice that to this order in $\alpha$ we can restore the integrations over $x,x\pm \muhat$ and we now recognize
that the right-hand side  of this expression
contains the propagator at the displaced points $F(x\pm 2\muhat).$\footnote{One might have imagined that
there are additional contributions arising from sites $x \pm \muhat \pm \nuhat$ but these in fact cancel
due to the staggered fermion phases.} A closed-form
expression for the latter can hence be found by going to momentum space where
\begin{equation}
F(p)=\frac{(i/\alpha) \sum_\mu\sin{p_\mu}}{\sum_\mu \sin^2{p_\mu}+m_F^2}
\end{equation}
with $m_F^2=-2+\frac{4}{\alpha^4}$.
Thus the strong-coupling calculation indicates that for sufficiently large $G$
the system should realize a phase in which the fermions acquire a mass without breaking the $SU(4)$ symmetry.

An analogous calculation can be performed for the bosonic propagator $B(x)=\vev{b(x)b(0)}$ corresponding
to the single-site fermion bilinear
$b=\psi^1\psi^2+\psi^3\psi^4$:
\begin{equation}
B(x)=2\delta_{x0}+\left(\frac{\alpha}{2}\right)^2\sum_\mu\left(B(x+\muhat)+B(x-\muhat)\right)
\end{equation}
or in momentum space
\begin{equation}
B(p)=\frac{8/\alpha^2}{4\sum_\mu\sin^2{p_\mu/2}+m_B^2}
\end{equation}
yielding a corresponding boson mass $m_B^2=-8+\frac{4}{\alpha^2}$. Thus one expects both bosonic and fermionic excitations to be gapped
at strong coupling. Furthermore,
this strong-coupling expansion suggests that the mechanism of dynamical mass generation in this model
corresponds to the condensation of a bilinear formed from the
original elementary fermions $\psi^a$ and a composite three-fermion state $\Psi^a=\epsilon_{abcd}\psi^b\psi^c\psi^d$ that
transforms in the complex conjugate representation of the $SU(4)$ symmetry. Clearly this is a non-perturbative phenomenon invisible in
weak-coupling perturbation theory.

The weak- and strong-coupling phases must be separated by at least one phase transition.
Previous work with similar lattice Higgs--Yukawa models employing staggered or naive
fermions had revealed such a paramagnetic strong-coupling (PMS) phase in a variety of models. However
such studies also typically revealed the presence of a third, intermediate phase in which the symmetries of the system
were spontaneously broken by the formation of a bilinear fermion condensate~\cite{Stephenson:1988td, Hasenfratz:1988vc, Lee:1989xq}.
In these earlier studies this
intermediate phase was separated from the weak- and strong-coupling regimes by first-order
phase transitions. One of the goals of the current work is to ascertain whether
such bilinear condensates appear at intermediate coupling in the current
model.

\section{\label{sec:aux}Auxiliary field representation}

We follow the standard strategy and rewrite the
original action (Eq.~\ref{action}) in a new form quadratic in the fermions but including an
auxiliary real scalar field. In our case this auxiliary field $\sigma^+_{ab}$ is an antisymmetric matrix in the internal space and possesses an important self-dual
property as described below.
This transformation preserves the free energy up to a constant:
\begin{equation}
S=\sum_{x,\mu}\psi^a \left[\eta.\Delta\,\delta_{ab}+G\sigma_{ab}^+\right]\psi^b+\frac{1}{4}\left(\sigma_{ab}^+\right)^2
\end{equation}
where
\begin{equation}\sigma_{ab}^+=P_{abcd}^+\sigma_{cd}=
\frac{1}{2}\left(\sigma_{ab}+\frac{1}{2}\epsilon_{abcd}\sigma_{cd}\right)\end{equation}
and we have introduced the projectors
\begin{equation}
P_{abcd}^\pm=\frac{1}{2}\left(\delta_{ac}\delta_{bd}\pm\frac{1}{2}\epsilon_{abcd}\right).\end{equation}
In principle one can now integrate over the fermions to produce a Pfaffian $\text{Pf}(M)$ where the
fermion operator $M$ is given by
\begin{equation}
M=\eta .\Delta  +G\sigma^+.\end{equation}
Rather remarkably one can show that the Pfaffian of this operator is in fact positive semi-definite. To see this
consider the associated eigenvalue equation
\begin{equation}
\left( \eta .\Delta+G\sigma^+\right)\psi=\lambda \psi.
\end{equation}
Since the operator is real and antisymmetric the eigenvalues of $M$ are pure imaginary and
come in pairs $i\lambda$ and $-i\lambda$. Sign changes in the Pfaffian would then
correspond to an odd number of eigenvalues
passing through the origin as the field $\sigma^+$ varies. But in our case we can
show that all eigenvalues are doubly degenerate, so no sign change is possible.

This degeneracy stems from the fact that $M$ is invariant under a set of $SU(2)$ transformations that form a subgroup
of the $SO(4)$ symmetry of the auxiliary field representation
with $SO(4) \simeq SU(2) \times SU(2)$. While the fermion transforms as a doublet under each of these
$SU(2)$s the auxiliary $\sigma^+$ is a singlet under one of them.\footnote{$\sigma^-$ is a singlet under the other $SU(2)$---this is just the standard representation theory of $SO(4)$.} Since the fermion operator is invariant under this
$SU(2)$ its eigenvalues are doubly degenerate. This conclusion has been
checked numerically and guarantees positivity of the Pfaffian. It is of crucial importance for our later
numerical work since it is equivalent to the statement that the system does not suffer from a sign problem---we
can replace $\text{Pf}(M)\to \det^{\frac{1}{4}}\left(MM^\dagger\right)$.

\section{\label{sec:phases}Phase diagram}

To probe the phase structure of the theory we first examine the square of the auxiliary
field $\frac{1}{4}\sigma_+^2=\frac{1}{2}\sum_{a<b}\left(\sigma_+^{ab}\right)^2$, which serves as a proxy for a four-fermion condensate and can
be computed analytically in the limits $G\to 0$ and $G\to\infty$. Consider the
modified action
\begin{equation}
  S\left(G,\beta\right) = \sum \frac{\beta}{4}\sigma_+^2+\sum\psi\left(\eta .\Delta+G\sigma_+\right)\psi.
\end{equation}
Clearly
\begin{equation}
  \vev{\frac{1}{4}\sigma_+^2} = -\frac{1}{V} \frac{\partial \ln Z\left(G, \beta\right)}{\partial\beta}.
\end{equation}
Rescaling $\sigma_+$ by $1/\sqrt{\beta}$ allows us to write the partition function $Z\left(G,\beta\right)$ as
\begin{equation*}
  Z\left(G, \beta\right) = \int D\sigma_+ \int D\psi \, e^{-S} = \beta^{-3V/2} Z\left(\frac{G}{\sqrt{\beta}}, 1\right)
\end{equation*}
where we have exploited the
antisymmetric self-dual character of $\sigma_+$ by allowing for just $3$ independent $\sigma$ integrations at each lattice site.
Thus
\begin{equation}
  \vev{\frac{1}{4}\sigma_+^2} = \frac{3}{2\beta} - \frac{1}{V} \frac{\partial \ln Z\left(\frac{G}{\sqrt{\beta}},1\right)}{\partial\beta}.
\end{equation}
Integrating over the fermions yields
\begin{equation}
Z\left(\frac{G}{\sqrt{\beta}},1\right)=\int D\sigma_+ \, \text{Pf}\left(\eta . \Delta+\frac{G}{\sqrt{\beta}}\sigma_+\right) e^{-\frac{1}{4}\sigma_+^2}.\end{equation}
For $G=0$ the partition function is $\beta$ independent, while its $\beta$ dependence is simply $\beta^{-V}$ in the strong-coupling
limit (Eq.~\ref{eq:strong}). Using these results and setting $\beta = 1$ gives
\begin{equation}
  \label{eq:sigma_vev}
  \vev{\frac{1}{4}\sigma_+^2} = \left\{\begin{array}{l} 3/2 \quad \mbox{as} \;\; G\to0 \\
                                                        5/2 \quad \mbox{as} \;\; G\to\infty. \end{array}\right.
\end{equation}
In practice we simulate the full antisymmetric $\sigma$ field  which allows us to monitor the vev of the
anti-selfdual component $\sigma_-$ also. Since this component does not couple to the fermions we expect $\vev{\frac{1}{4}\sigma_-^2}=3/2$ independent of $G$.

\begin{figure}[btp]
  \centering
  \includegraphics[width=\linewidth]{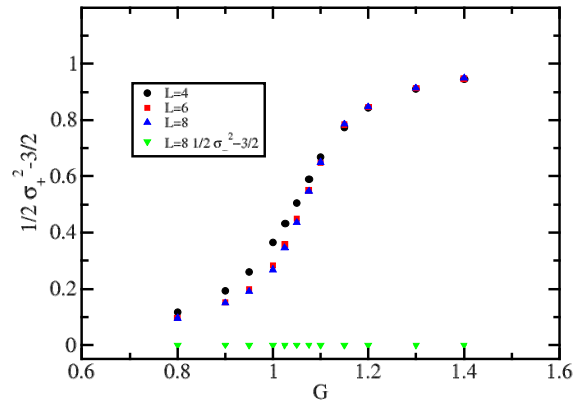}
  \caption{\label{plus}$\vev{\frac{1}{4}\sigma_{\pm}^2} -\frac{3}{2}$ vs.~$G$ for $L=4$, 6 and 8 with vanishing external sources ($m=0$ in Eq.~\protect\ref{eq:sources}).}
\end{figure}

Our numerical
results for $\vev{\frac{1}{4}\sigma_{\pm}^2} -\frac{3}{2}$ shown in Fig.~\ref{plus} are consistent with these predictions. The observed behavior of $\sigma_+^2$ appears to interpolate smoothly between the weak- and strong-coupling limits of Eq.~\ref{eq:sigma_vev}, while $\sigma_-^2$ shows no dependence on $G$ as
expected.  There
are no signs of first-order phase transitions and indeed on $L^4$ lattices with $L > 4$ the observed finite-volume effects are small.
In our simulations we have employed thermal boundary conditions: the fermions wrapping the
temporal direction pick up a minus sign. This has the merit of removing an exact fermion zero mode arising at $G=0$
and preserves all symmetries of the system.\footnote{This corrects a comment in our earlier
paper~\cite{Catterall:2015zua}, which stated that thermal boundary conditions break the shift symmetries.  We thank Shailesh Chandrasekharan for the correction.}

\begin{figure}[btp]
  \centering
  \includegraphics[width=\linewidth]{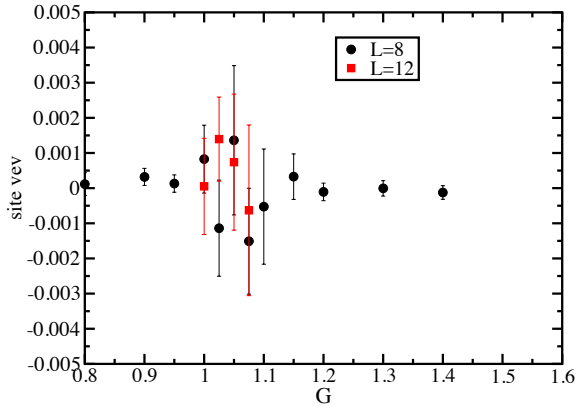}
  \caption{\label{dis} Site bilinear vs.~$G$ for $L = 8$ and 12 with zero external sources.}
\end{figure}

The transition from weakly coupled free fields to strongly coupled four-fermion condensates is most
clearly seen by plotting a susceptibility defined by
\begin{equation}
\chi=\frac{1}{V}\sum_{x,y,a,b}
\left\langle \psi^a(x)\psi^b(x)\psi^a(y)\psi^b(y)\right\rangle.
\end{equation}
Using Wick's theorem this can be written as sums of products of fermion propagators. We group these
into connected and disconnected contributions
\begin{align}
  \chi_{\text{conn}} & = \frac{1}{V}\sum_{x,y} \Big[\Big\langle \psi^a(x)\psi^a(y)\Big\rangle \Big\langle \psi^b(x)\psi^b(y)\Big\rangle          \\
                  & \hspace{50 pt}             - \Big\langle \psi^a(x)\psi^b(y)\Big\rangle \Big\langle \psi^b(x)\psi^a(y)\Big\rangle\Big] \nn
\end{align}
\begin{equation}
  \chi_{\text{dis}} = \frac{1}{V}\left[\sum_x \vev{\psi^a(x)\psi^b(x)}\right]^2,
\end{equation}
respectively.
The disconnected contribution $\chi_{\text{dis}}$ should vanish by symmetry in finite volume, and we have verified that this is indeed the case: See Fig.~\ref{dis} in which we plot the bilinear expectation value that is responsible for $\chi_{\text{dis}}$.
As expected it is statistically consistent with zero for all values of the four-fermi coupling.
If one assumes a non-zero vev consistent with the error bars one can easily see that the corresponding disconnected susceptibility $\chi_{\text{dis}} < 0.1$ for all $G$.

\begin{figure}[btp]
  \centering
  \includegraphics[width=\linewidth]{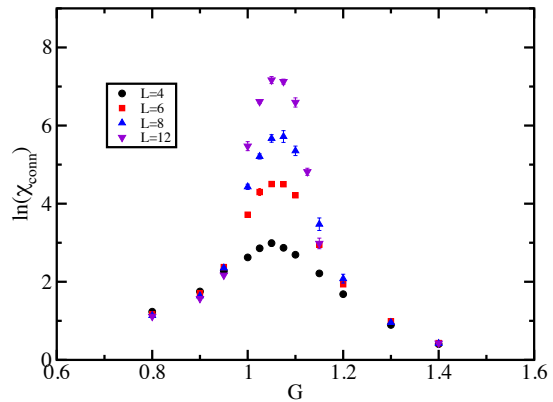}
  \caption{\label{lsus}$\ln{\chi_{\text{conn}}}$ vs.~$G$ for $L=4$, 6, 8 and 12 with zero external sources.}
\end{figure}

This is much smaller than the connected contribution $\chi_{\text{conn}}$, the logarithm of which we plot in Fig.~\ref{lsus}.
The fermion propagators used in this measurement were obtained by inverting the fermion operator on sixteen
point sources located at $(p_1,p_2,p_3,p_4)$ with $p_i\in \left\{0,L/2\right\}$ on each configuration and subsequently averaging
the results over the Monte Carlo ensemble. A well-defined peak that scales rapidly with increasing
volume is seen centered around $G_c \approx 1.05$. The position, width and height of this peak
agree well with those reported in \cite{Ayyar:2016lxq}, using the
mapping $G^2=\frac{2}{3}U$ to relate our coupling $G$ to the coupling $U$ appearing in that work. This mapping requires rescaling
the fermions by a factor of $\sqrt{2}$ to fix the coefficient of the kinetic term.

\begin{figure}[btp]
  \centering
  \includegraphics[width=\linewidth]{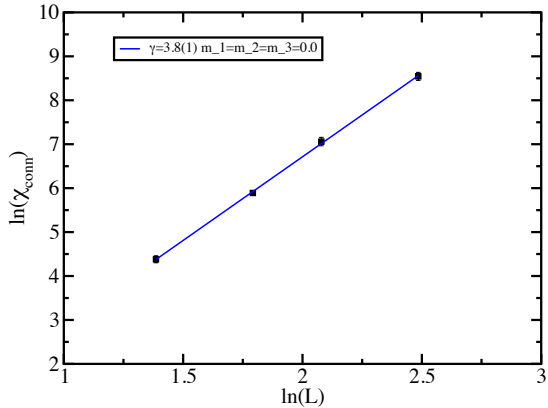}
  \caption{\label{lpeak}$\ln{\chi_{\text{conn}}}$ vs.~$\ln{L}$ at $G=1.05$ for zero external sources.  A least-squares fit to the power law $\chi_{\text{conn}} \propto L^{\gamma}$ yields $\gamma = 3.8(1)$.}
\end{figure}

If we assume that the height of the connected susceptibility peak scales as $\chi_{\text{max}}\sim L^\gamma$ we can estimate $\gamma$ from a log--log plot
of the susceptibility versus the lattice size. Such a plot is shown in Fig.~\ref{lpeak}.
The value $\gamma = 3.8(1)$ extracted from a fit is in approximate agreement with the volume scaling reported
in~\cite{Ayyar:2016lxq} for the {\it full} susceptibility $\chi$.  In the latter work the volume
scaling is attributed to the formation of
an $SU(4)$-breaking fermion bilinear condensate.
However, such a condensate would be associated with the disconnected contribution $\chi_{\text{dis}}$ which is {\it not} included in Fig.~\ref{lsus}.  We conclude that
whatever is the reason for the volume scaling of the susceptibility $\chi$ it does
not require the appearance of a bilinear fermion condensate.  Indeed, in the following section we
have looked carefully for the appearance of such a condensate and see no evidence for it.

\begin{figure}[btp]
  \centering
  \includegraphics[width=\linewidth]{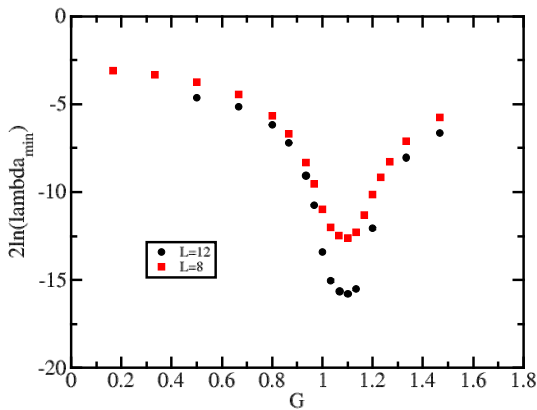}
  \caption{\label{eigen}$2\ln{\lambda_{\text{min}}}$ vs.~$G$ for $L = 8$ and 12 with zero external sources.}
\end{figure}

Instead to explain the divergence of the connected susceptibility the system must
develop long-range correlations.  One piece of evidence for this can be seen in Fig.~\ref{eigen} where we plot the logarithm of the smallest eigenvalue of the fermion
operator vs.~the four-fermi coupling.
The smallest eigenvalue
falls rapidly in a region between $G\approx1.0$--1.1 consistent with the peak seen in the
connected susceptibility.\footnote{This dramatic drop in the smallest eigenvalue
is paired with a corresponding rapid increase in the number of conjugate gradient (CG) iterations
needed to invert the fermion operator. It is this fact that has limited the largest lattice
that we can easily simulate; at the critical point with zero external sources the $L=12$ lattice requires approximately 20,000
CG iterations per solve.}

\begin{figure}[btp]
  \centering
  \includegraphics[width=\linewidth]{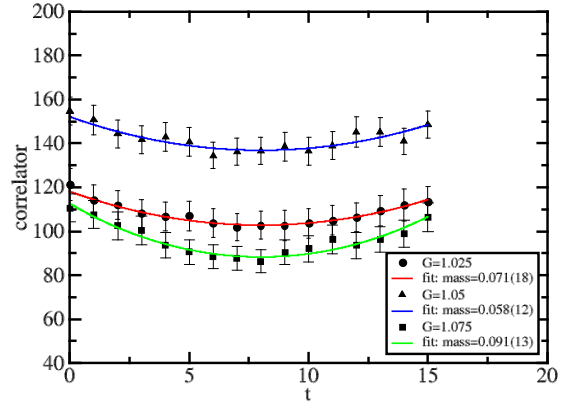}
  \caption{\label{corr}Timeslice-averaged correlator $G(t)$ of bilinear density for several couplings $G$ around the critical region, on $8^3\times 16$ lattices with zero external sources.  The lines are cosh fits.}
\end{figure}

We can gain further insight into this issue by computing the bosonic two-point function whose temporal
sum yields $\chi_{\text{conn}}$:
\begin{equation}
\chi_{\text{conn}}=\frac{1}{V}\sum_t G(t)
\end{equation}
where
\begin{align}
    G(t) & = \frac{1}{V}\sum_{x,y,a,b}\Big(\Big\langle \psi^a(x)\psi^a(y)\Big\rangle \Big\langle \psi^b(x)\psi^b(y)\Big\rangle \\
         & \hspace{27 pt}                - \Big\langle \psi^a(x)\psi^b(y)\Big\rangle \Big\langle \psi^b(x)\psi^a(y)\Big\rangle\Big)\delta(x_t - y_t - t) \nn
\end{align}
and the $\delta$ function picks out points separated by
$t$ units in the time direction.
This connected correlation function $G(t)$
is shown in Fig.~\ref{corr} for $8^3\times 16$ lattices.
The solid lines are cosh fits and allow us to read off the mass of the bosonic state
created by operating on the vacuum with the bilinear operator $\psi^a(x)\psi^b(x)$.

\begin{figure}[btp]
  \centering
  \includegraphics[width=\linewidth]{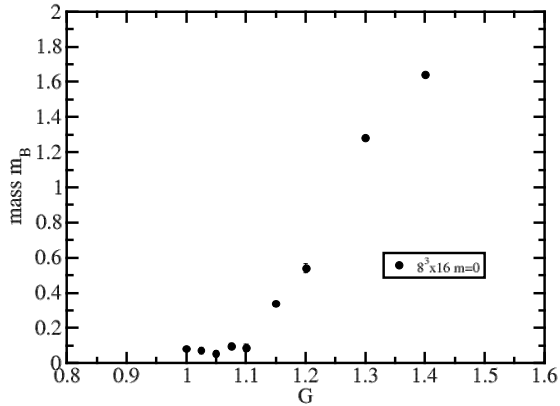}
  \caption{\label{mass}Mass of the bilinear state $B^{ab} = \psi^a\psi^b$ versus $G$, for $8^3\times 16$ lattices with zero external sources.  Most error bars are smaller than the symbols.}
\end{figure}

Figure~\ref{mass} shows
this mass as a function of the coupling $G$.  At strong coupling the
mass rises quickly as expected from the strong-coupling expansion.  But in the critical region $1.0 \leq G \leq 1.1$ corresponding to the peak in
the susceptibility the mass is very small and independent of $G$.
This structure together
with the observed rather broad peak in the susceptibility prompts one to conjecture that the system may
indeed possess a narrow intermediate phase as
reported in~\cite{Ayyar:2016lxq}.
Where we differ from~\cite{Ayyar:2016lxq} is in the question of
whether such a phase is characterized by a bilinear condensate.
In the next section we study the model with external symmetry-breaking
source terms and find no evidence of a fermion condensate formed from either single-site
or multilink bilinear operators.

\section{\label{sec:cond}Bilinear condensates and spontaneous symmetry breaking}

To probe the question of spontaneous symmetry breaking, we have augmented the
action shown in Eq.~\ref{action} by adding source terms which couple to both $SU(4)$-breaking fermion bilinear terms
and the shift-symmetry-breaking one-link terms described in Eq.~\ref{ops}:
\begin{equation}
  \label{eq:sources}
  \begin{split}
    \Delta S = & \sum_{x,a,b} \left(m_1+\epsilon(x)m_2\right) \left[\psi^a(x)\psi^b(x)\right]_+\Sigma^{ab} \\
               & + m_3\sum_{x,\mu,a} \epsilon(x)\xi_\mu(x)\psi^a(x)S_\mu\psi^a(x),
  \end{split}
\end{equation}
where we choose the $SU(4)$-breaking source term
\begin{equation}
  \label{Sigma}
  \Sigma^{ab}=\left(\begin{array}{cc} i\sigma_2 & 0 \\ 0 & i\sigma_2\end{array}\right).
\end{equation}
Notice that we allow for both a regular and staggered single-site fermion bilinear in this expression.  The
latter operator breaks all the exact symmetries of the
action but appears as a rather
natural mass term when the model is rewritten in terms of two {\it full} staggered fields.\footnote{We
thank Shailesh Chandrasekharan for pointing this out.}

\begin{figure*}[btp]
  \includegraphics[width=0.45\linewidth]{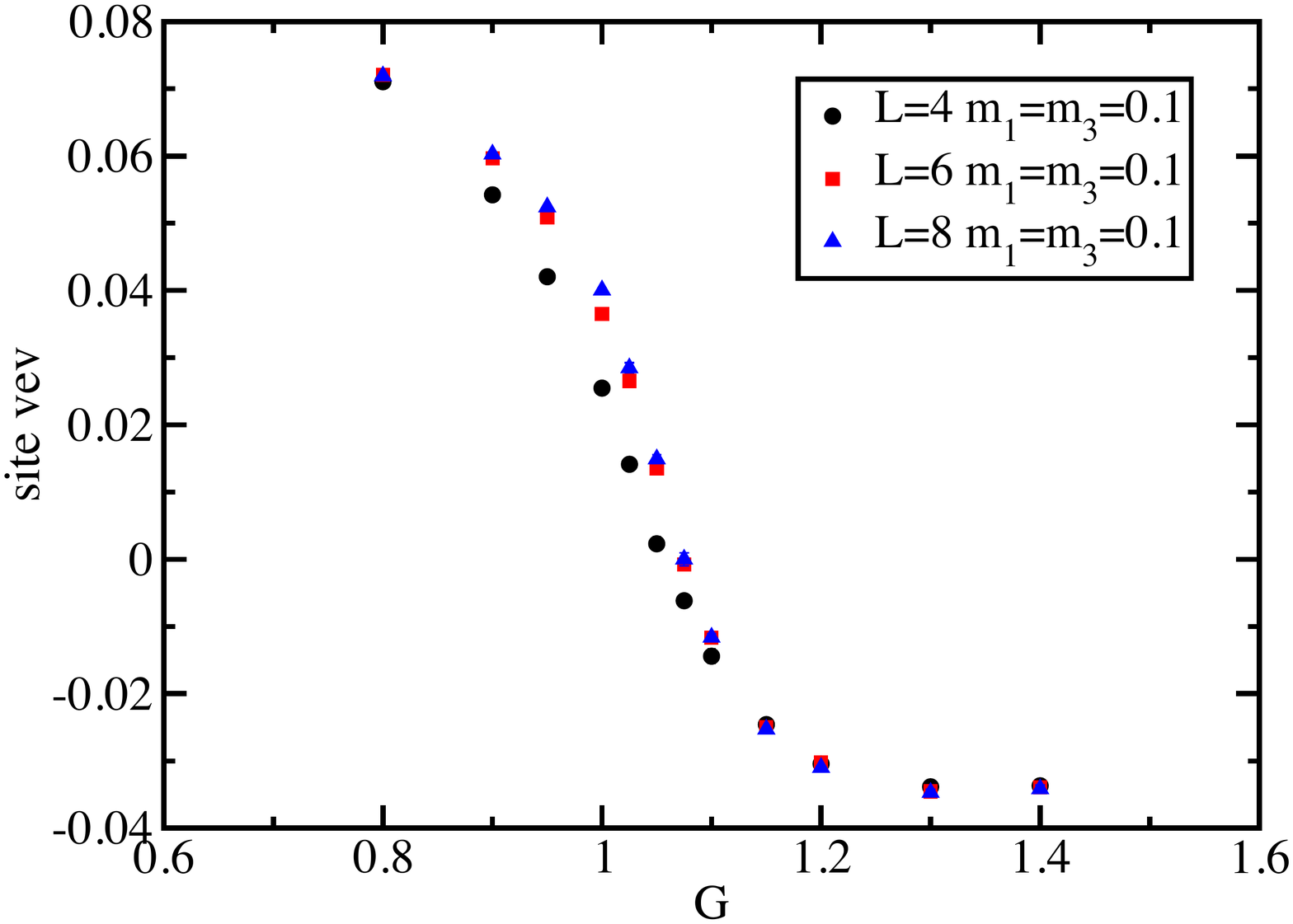} \hfill \includegraphics[width=0.45\linewidth]{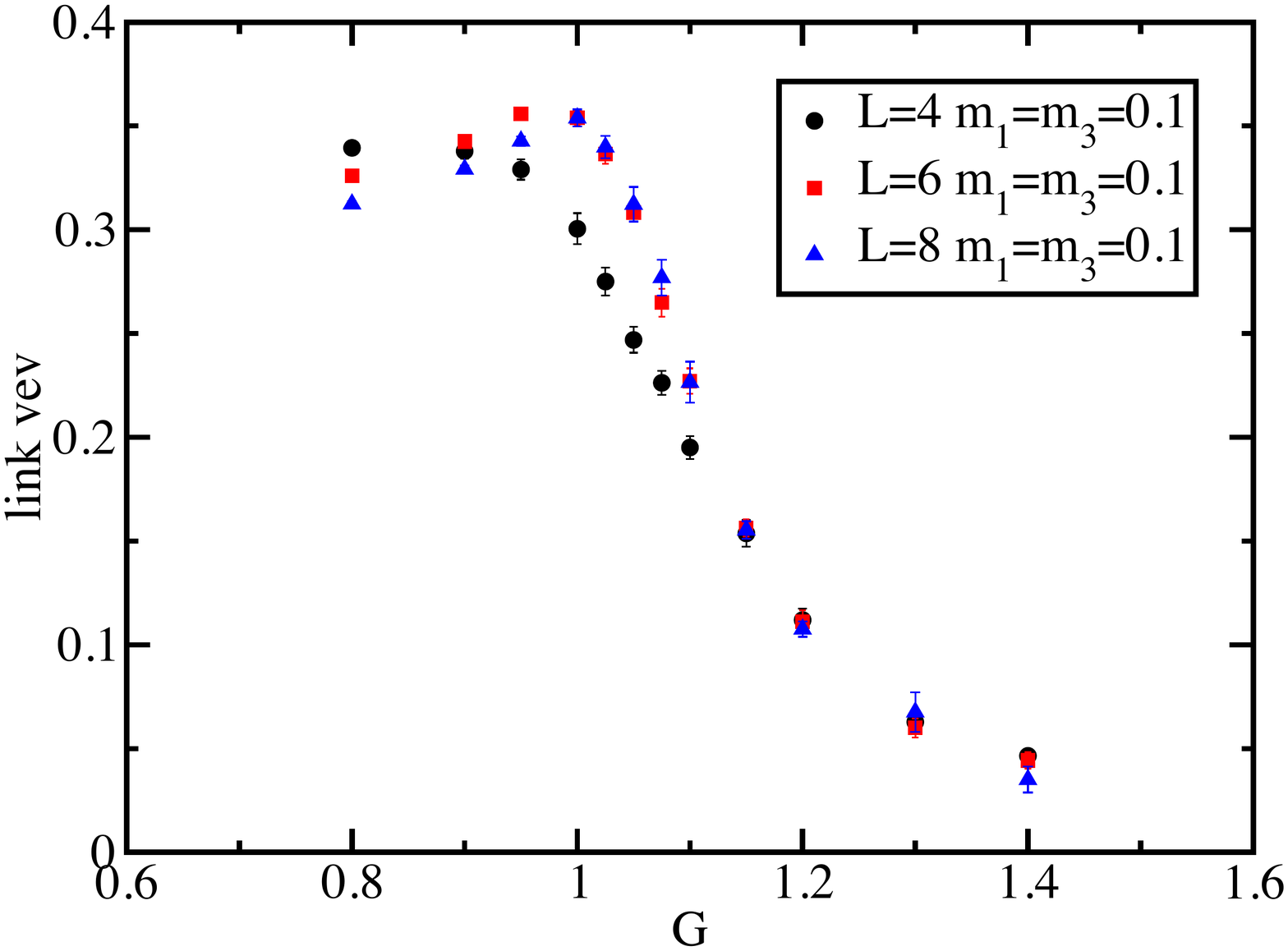}
  \caption{\label{pscan}Site (left) and link (right) bilinears vs.~$G$ for $L=4$, 6 and 8 with external source couplings $m_1=m_3=0.1$ and $m_2=0$.}
\end{figure*}

We have additionally assumed a rotationally invariant form of the
coupling to the one-link term. The results for the link and site bilinear vevs from runs with $m_1=m_3=0.1$ and $m_2=0$ with varying $G$ are shown
in Fig.~\ref{pscan}.
While the presence of the source terms clearly leads to non-zero vevs for the bilinears at any coupling $G$, these plots make it clear that these vevs are monotonically suppressed as one enters the strongly coupled regime.
Of course to look for symmetry breaking we should fix the four-fermi coupling and examine the behavior of these vevs in the thermodynamic limit as the external source is sent to zero.
Since any would-be symmetry breaking must occur in the critical regime $1.0 \leq G \leq 1.1$ we initially fix $G=1.05$ while varying the external sources.

\begin{figure*}[btp]
  \includegraphics[width=0.45\linewidth]{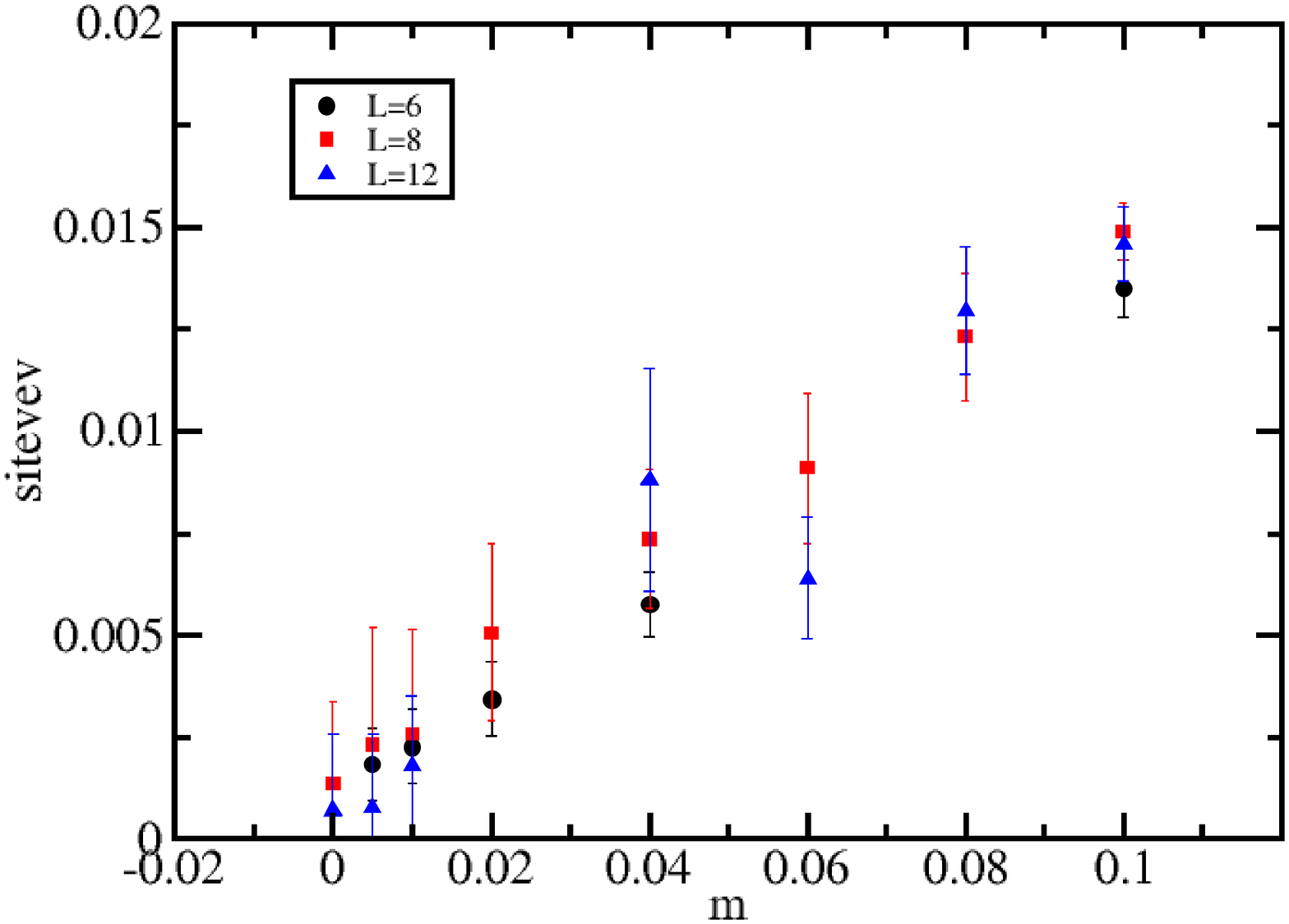} \hfill \includegraphics[width=0.45\linewidth]{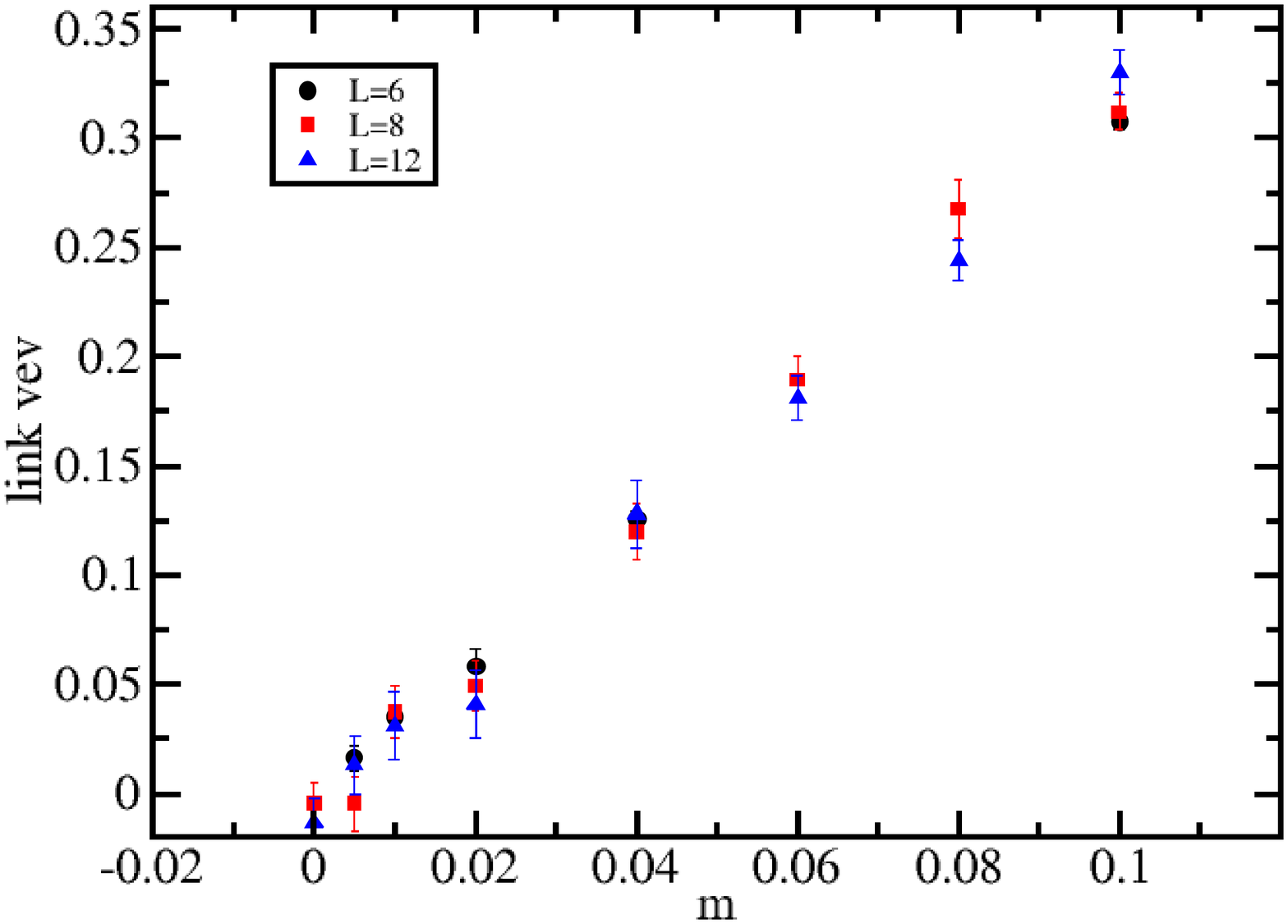}
  \caption{\label{bilinears1}Site (left) and link (right) bilinears vs.~$m$ for $L=6$, 8 and 12 at $G=1.05$ with external source couplings $m_1=m_3=m$ and $m_2=0$.}
\end{figure*}

The results of such a study are shown in Fig.~\ref{bilinears1} for $G=1.05$, $m_1=m_3=m$ and $m_2=0$.
As expected the vevs vanish on any finite-volume system in the limit in which the external field is sent to zero as a consequence of the exact lattice symmetries which appear in that limit.
A signal of spontaneous symmetry breaking would be a condensate that
grows with volume for small enough values of the external source.  Such behavior would allow for the
possibility that the condensate remains finite in the thermodynamic limit as the source is removed.
This occurs, for example, in the reduced staggered four-fermion model studied by Ref.~\cite{Catterall:2013sto}, where the signal for spontaneous symmetry breaking via the formation of a bilinear is very clear even on small lattices.

\begin{figure*}[btp]
  \includegraphics[width=0.45\linewidth]{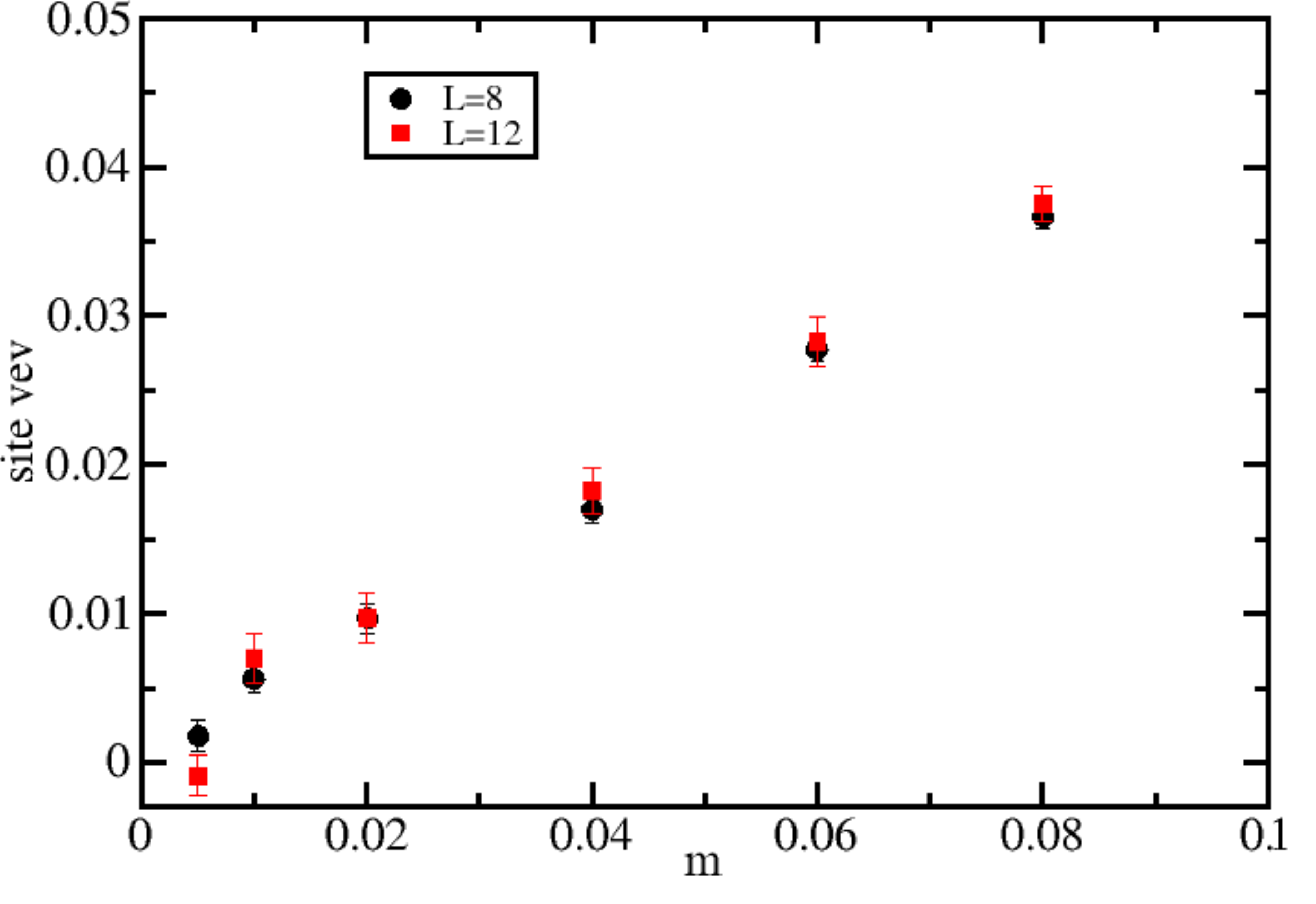} \hfill \includegraphics[width=0.45\linewidth]{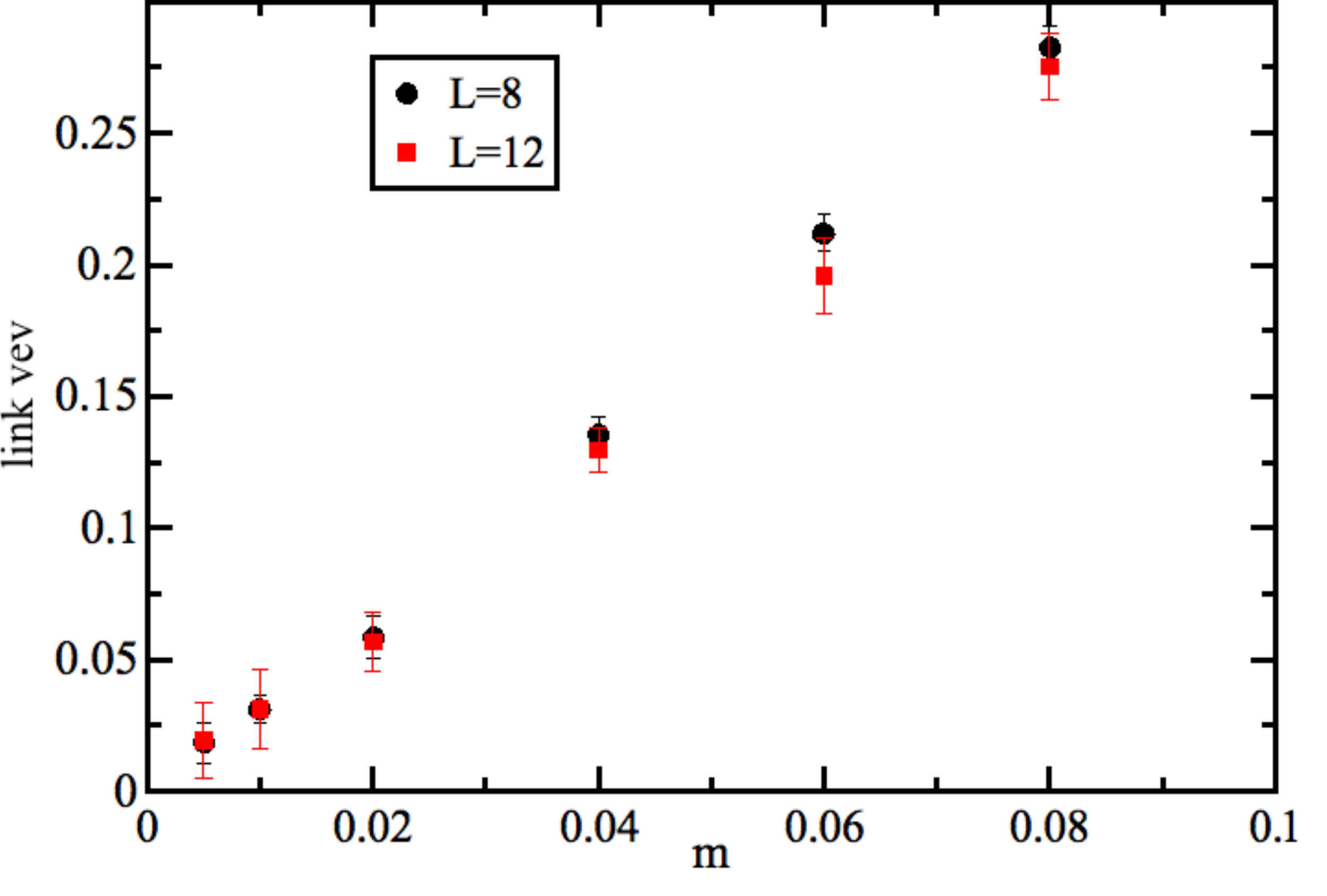}
  \caption{\label{old}Site (left) and link (right) bilinears vs.~$m$ for $L=8$ and 12 at $G=1.0$ with external source couplings $m_1=m_3=m$ and $m_2=0$.}
\end{figure*}

The results shown in Fig.~\ref{bilinears1} are {\it not} consistent with this scenario: the finite-volume effects
are small for both the single-site bilinear and the link bilinear for small external sources.
We conclude that our numerical results for these particular bilinear terms are not compatible with
spontaneous breaking of either the shift or $SU(4)$ symmetries.
This conclusion extends to all couplings $G<1.05$, as illustrated by Fig.~\ref{old} for $G = 1$.
These results are strengthened by the calculation presented in Sec.~\ref{sec:CW}, which shows that the one-loop effective potential for the auxiliary field $\sigma_+$ retains a minimum at the origin for any value of $G$---a result consistent with the vanishing vev of the single-site bilinear examined here.

\begin{figure*}[btp]
  \includegraphics[width=0.45\linewidth]{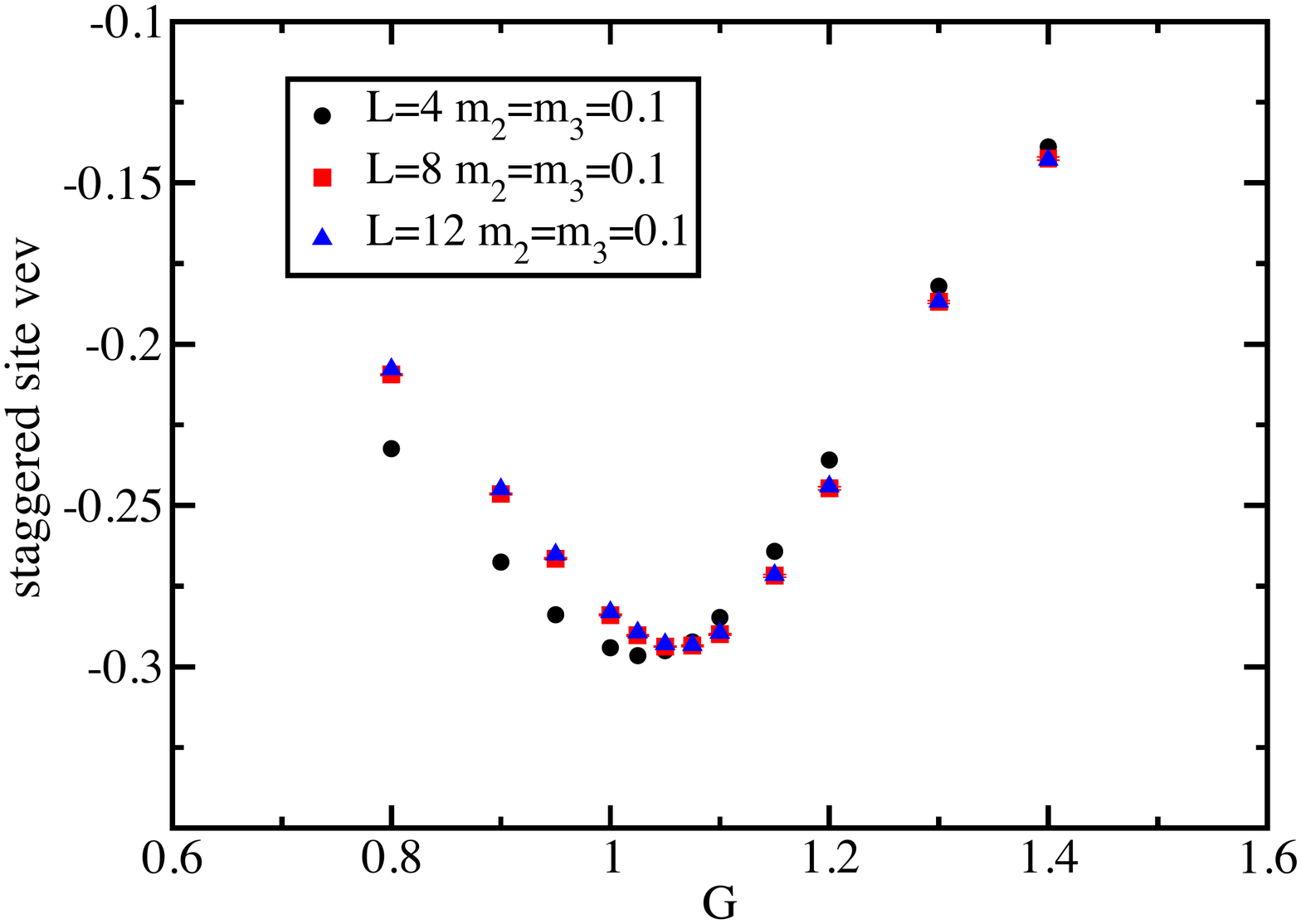} \hfill \includegraphics[width=0.45\linewidth]{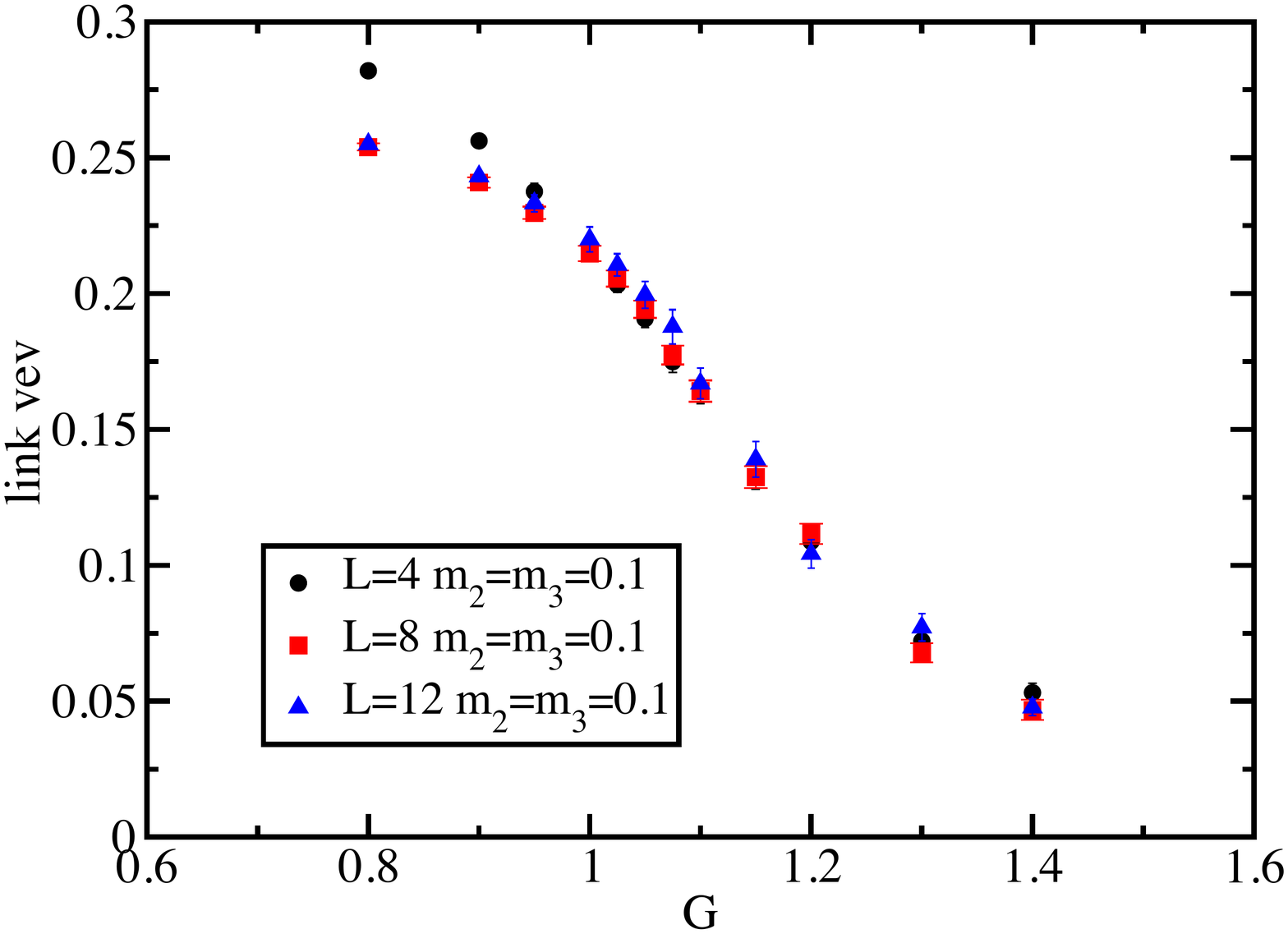}
  \caption{\label{bilinears}Staggered site (left) and link (right) bilinears vs.~$G$ for $L=4$, 8 and 12 with external source couplings $m_2=m_3=0.1$ and $m_1=0$.}
\end{figure*}

We have also examined the model in the presence of the staggered single-site bilinear term corresponding
to $m_2=m_3=0.1$ and $m_1=0$ and show the results in Fig.~\ref{bilinears}.
The vev of the link operator in Fig.~\ref{bilinears} is again driven monotonically to zero with increasing coupling $G$ but
the staggered site bilinear shows more interesting behavior---its magnitude attains a maximum precisely
in the critical regime $1.0 \leq G \leq 1.1$.  This suggests that in this region the system may be trying to
form a {\it staggered} bilinear condensate.  Such a staggered vev would be invisible
to an order parameter that simply averages over the lattice sites without regard to site parity, such as
the single-site bilinear examined above.  A non-zero staggered
vev would nevertheless correspond to $SU(4)$ symmetry breaking.

\begin{figure*}[btp]
  \includegraphics[width=0.45\linewidth]{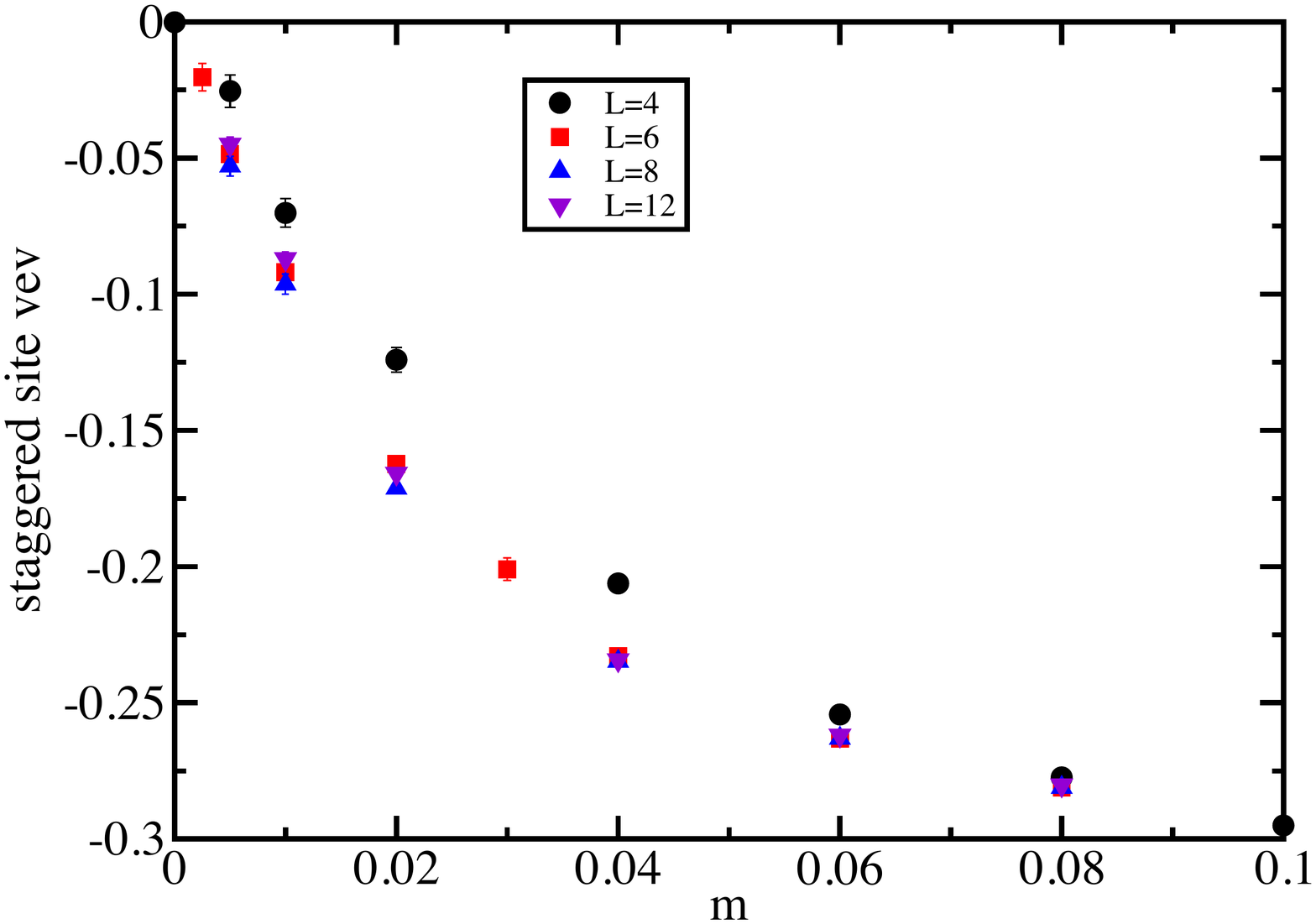} \hfill \includegraphics[width=0.45\linewidth]{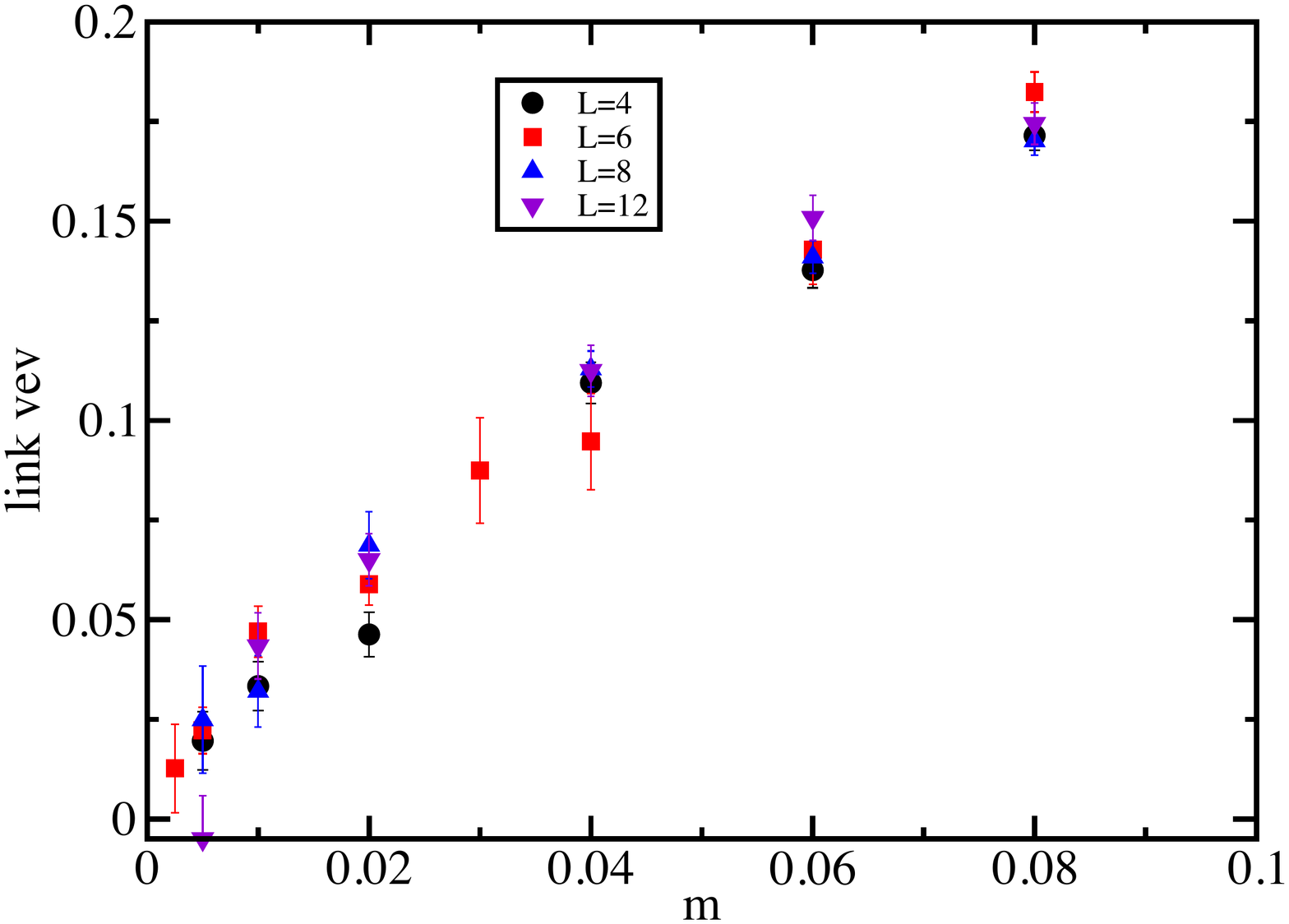}
  \caption{\label{stag}Staggered site (left) and link (right) bilinears vs.~$m$ for $L=4$, 6, 8 and 12 at $G=1.05$ with external source couplings $m_2=m_3=m$ and $m_1=0$.}

\end{figure*}
\begin{figure*}[btp]
  \includegraphics[width=0.45\linewidth]{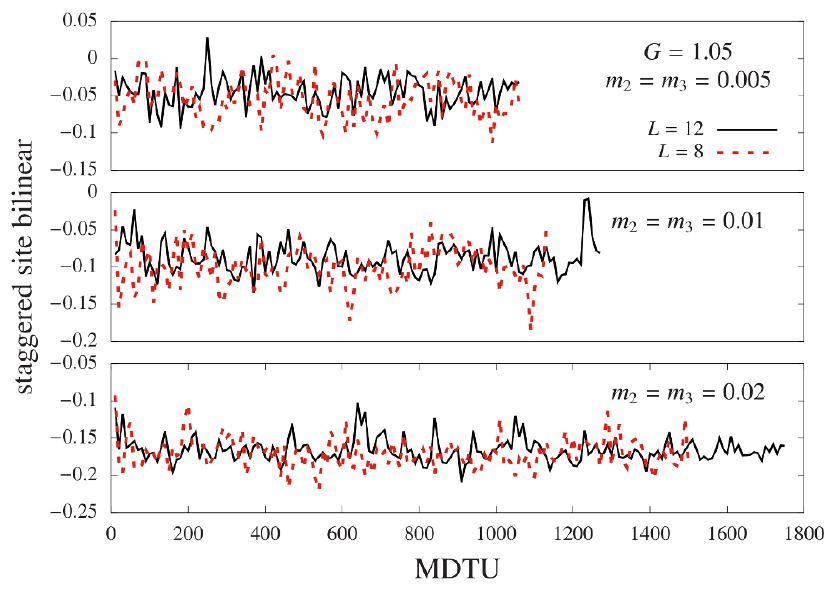} \hfill \includegraphics[width=0.45\linewidth]{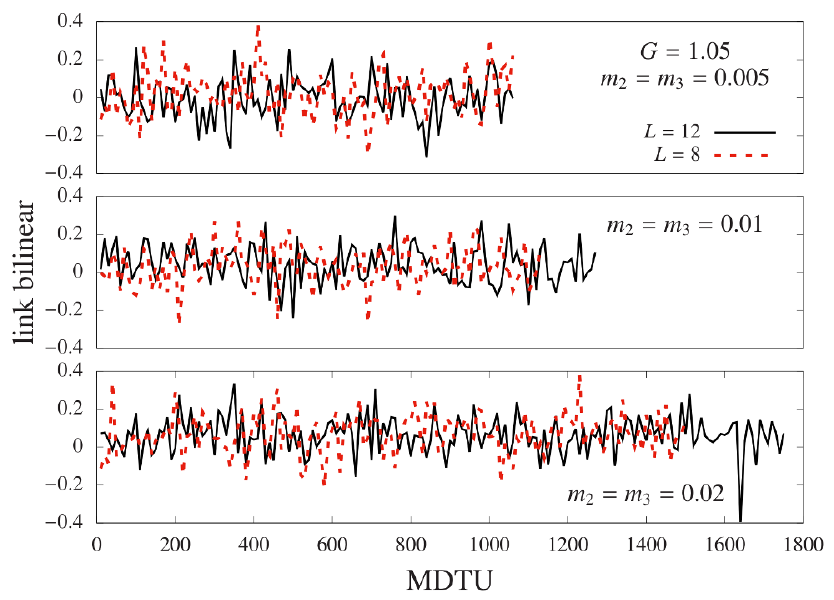}
  \caption{\label{time_series}Monte Carlo time-series plots for the staggered site (left) and link (right) bilinears for three small values of $m_2 = m_3 = 0.005$, 0.01 and 0.02 (from top to bottom) with $m_1 = 0$ and $G = 1.05$.  In each case there is no visible change between the data for $L = 8$ (dashed red lines) and $L = 12$ (solid black lines), reflecting the very weak volume dependence of the corresponding vevs shown in Fig.~\protect\ref{stag}.}
\end{figure*}

Again, to see whether such a symmetry breaking pattern occurs we
have examined the volume dependence of this staggered bilinear vev as a
function of the external source $m$.
The results are shown in Fig.~\ref{stag}.  Again the volume dependence for both the link and now the staggered
site bilinear is very weak and there is no sign that spontaneous symmetry breaking will occur in
the thermodynamic limit as the source is removed.
This conclusion is supported by the Monte Carlo time-series plots in Fig.~\ref{time_series}, which show representative raw data for several of the points with small $m \leq 0.02$ in Fig.~\ref{stag}.
For both the staggered site and link bilinear these time series show no visible change between the $L = 8$ data and that for $L = 12$.

\begin{figure}[btp]
  \centering
  \includegraphics[width=\linewidth]{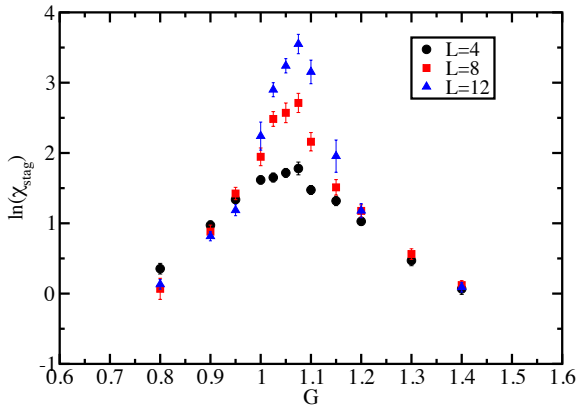}
  \caption{\label{stagsus}Staggered susceptibility vs.~$G$ for $L=4$, 8 and 12 with zero external sources.}
\end{figure}

To summarize we have examined three separate bilinear operators---the
single-site, staggered single-site and one-link operators---for signals of non-zero symmetry-breaking condensates and find a null result.  The staggered
single-site operator is interesting as it shows the strongest response to an external field but even
in this case there is no evidence that it forms a condensate in the critical region.  Nevertheless, it is interesting
to examine the corresponding staggered susceptibility
\begin{equation}
  \chi_{\text{stag}} = \frac{1}{V}\left(\vev{O_{\text{stag}}^2} - \vev{O_{\text{stag}}}^2\right)
\end{equation}
with $O_{\text{stag}} = \sum_x\epsilon(x)\left[\psi^0(x)\psi^1(x)\right]_+$.
This is shown in Fig.~\ref{stagsus} as a function of $G$ with no external
sources.
While this staggered susceptibility diverges in the same
critical regime as before it does so with a significantly smaller exponent
than the susceptibility considered earlier.  A least-squares fit to $\chi_{\text{stag}}\sim L^p$ yields an exponent $p=1.55(14)$ with a $\chi^2/\text{dof}=1.2$.  Such an
exponent would correspond to a continuous transition and
yields a scaling dimension $\Delta \sim 1.2$ for the staggered bilinear.
Of course confidence in the value of this scaling exponent will require the use of larger lattices than those employed in the current study.
This is underway.

\section{\label{sec:CW}Coleman--Weinberg effective potential}
One standard way to look for spontaneous symmetry breaking in four-fermi theories is to compute the one-loop effective potential for the $\sigma_+$ field.\footnote{We thank Jan Smit for pointing out this possibility.}
After integrating over the fermions the effective action takes the form
\begin{equation}
  S_{\text{eff}}(\sigma_+) = -\frac{1}{2} \text{Tr}\ln{\left(\eta .\Delta+G\sigma_+\right)}.
\end{equation}
In a constant $\sigma_+ = \mu\Sigma$ background (Eq.~\ref{Sigma}) we can diagonalize the kinetic operator and exploit its real antisymmetric character to derive the effective potential
\begin{equation*}
  V_{\text{eff}}(\mu) = -\frac{1}{4} \text{tr} \sum_k \left(\ln\left[i\lambda_k + G\mu\Sigma\right] + \ln\left[-i\lambda_k + G\mu\Sigma\right]\right)
\end{equation*}
where $\text{tr}$ denotes the remaining trace over $SU(4)$ indices and $\pm i\lambda_k$ are eigenvalues of $\eta.\Delta$.
Collecting terms and carrying out the final trace yields
\begin{equation}
  \begin{split}
    V_{\text{eff}}(\mu) & = -\sum_k \ln\left[\lambda_k^2 - G^2\mu^2\right] \\
                        & = V(0) - \sum_k\ln\left[1 - \frac{G^2\mu^2}{\lambda_k^2}\right].
  \end{split}
\end{equation}
One can see that the effective potential is extremized at $\mu = 0$ and it is trivial to further show that
\begin{equation}
  \left.\frac{\partial^2 V_{\text{eff}}}{\partial\mu^2}\right|_{\mu = 0} > 0
\end{equation}
independent of $G$.
Thus the symmetric state $\mu = 0$ remains a local minimum of the effective potential and the vev of $\sigma_+$ vanishes for all $G$---there can be no spontaneous symmetry breaking at least in the one-loop approximation.

\section{\label{sec:concl}Conclusions}

In this paper we have studied perhaps the simplest relativistic lattice four-fermion model one can construct comprising exactly four Grassmann degrees of
freedom per site arising from four {\it reduced} staggered fermions.
We have argued that the system will possess a symmetric gapped phase for large four-fermi coupling but will describe eight free massless
Dirac fermions in the continuum limit at weak coupling.  For a narrow region of intermediate couplings we
have observed that the system develops long-range correlations.
In all the
earlier work on lattice Higgs--Yukawa theories the appearance of such long-range correlations was associated
with the appearance of an intermediate ferromagnetic
phase characterized by a symmetry-breaking fermion bilinear condensate.
These earlier works typically employed a scalar kinetic term with hopping parameter $\kappa$, and sometimes also a quartic scalar coupling $\lambda$, in addition to the scalar mass term and Yukawa interaction.
Thus the $\kappa = \lambda = 0$ line in these earlier phase diagrams would come closest to the model described here.
An example of such a phase diagram is Figure~1 in Ref.~\cite{Bock:1990cx}, which makes it clear that even along the line $\kappa = 0$ a ferromagnetic phase separates the PMW and PMS phases.

In the current study we have searched for the appearance of such bilinear symmetry-breaking condensates explicitly by sourcing the system with a variety of fermion
bilinear mass terms and examined carefully the response of the system as these source terms are sent to zero.  The results of our calculations are completely
consistent with the absence of bilinear condensates at all couplings.
We have strengthened this conclusion with an analytic calculation of the one-loop effective potential for the auxiliary field.
For this model we find that $V_{\text{eff}}$ is {\it not} of the symmetry-breaking form, in agreement with the numerical results.
Thus the current model appears to exhibit quite different behavior from those studied earlier.

One reason for the difference may be the different fermion discretizations used in the various studies.
The naive, Wilson or regular staggered fermions employed in the past enjoy a different set of
exact lattice symmetries, and in particular allow for symmetric single-site mass terms that are absent for the reduced staggered fermions used here.
Although Ref.~\cite{Bock:1992yr} also uses reduced staggered fermions,
it considers the limit $\lambda \to \infty$ rather than $\lambda = 0$ and employs a four-fermi term based on the square
of a one-link mass operator, which means that discrete shift symmetries rather than continuous lattice
symmetries are broken by the formation of a condensate.
Since the exact lattice symmetries are not the same, we cannot assume that the same critical behavior should be observed at non-zero four-fermi coupling.

While we see no signs of a broken phase we do see strong signs of critical behavior at intermediate coupling:
Susceptibilities associated with certain fermion four-point functions diverge with increasing lattice size in a narrow region of the four-fermi coupling and the mass of a certain composite boson
approaches zero.
In Ref.~\cite{Ayyar:2016lxq} the volume scaling of this
susceptibility was interpreted as evidence for a narrow intermediate phase with broken
$SU(4)$ symmetry.  This phase structure would necessarily imply the existence of
two phase transitions.  Our results are compatible with the appearance of a narrow intermediate phase, but
indicate that no symmetry-breaking bilinear condensate forms in this regime.  Given the absence of
an obvious local order parameter we remain agnostic as to whether the phase diagram
contains a narrow intermediate phase or a single phase transition directly separating the weak- and strong-coupling regimes.
We plan further studies to test these two possibilities.

The observed phase structure
is somewhat reminiscent of the two-dimensional Thirring model which develops
a mass gap without breaking chiral symmetry~\cite{Witten:1978qu}.\footnote{We thank Simon Hands for bringing this and related papers to our attention~\cite{Hands:2001cs}.} In the
two-dimensional case the corresponding susceptibility is the integral of the four-point function which develops power-law scaling for
strong coupling,
\begin{equation}
  \vev{\overline{\psi}(0)\psi(0)\overline{\psi}(r)\psi(r)} \sim \frac{1}{r^x}
\end{equation}
where $x\sim 1/N_f$ and $N_f$ is the number of continuum flavors.  This model also possesses
a phase transition without an order parameter,
driven by the condensation of topological defects associated with the auxiliary field introduced to represent the effects of the four-fermi interaction.
Of course the physics in two dimensions is quite different from four dimensions so one must be careful in pursuing this analogy too far.
Even so, we note that the would-be breaking pattern $SU(2) \to U(1)$ does allow for topological field configurations---Hopf defects---to exist in the four-dimensional model.

There has been considerable interest in recent years within the condensed matter community in the
construction of models in which fermions can be gapped without breaking symmetries
using carefully chosen quartic interactions~\cite{Fidkowski:2009dba, Morimoto:2015lua}. Although the condensed matter
models are constructed using Hamiltonian
language and describe non-relativistic
fermions it is nevertheless intriguing that the sixteen Majorana fermions they require match the sixteen
Majorana fermions that are expected at weak coupling in this lattice theory.
It has been proposed that such quartic interactions can be used in the context of domain wall fermion theories to provide a path to achieve chiral lattice gauge theories~\cite{You:2014vea, BenTov:2015gra, Wang:2013yta}. If indeed the current model avoids
symmetry-breaking phases it may be possible to revisit the
original Eichten--Preskill proposal for the construction of chiral lattice gauge theories using strong four-fermion
terms in the bulk to lift fermion doubler modes~\cite{Eichten:1985ft, Poppitz:2010at}. However,
it is not clear to the authors how such constructions can work in detail; the model
described here uses reduced staggered rather
than Wilson or naive fermions which negates a simple transcription of the four-fermion interaction appearing in this model to those earlier constructions.

Independently of these speculations one can wonder whether the phase transition(s) in the model described here
are evidence of new continuum limit(s) for strongly interacting fermions in four dimensions. One must
be careful in drawing too strong a conclusion at this stage; even if a new fixed point exists it might
not be Lorentz invariant. Indeed, given the connection between staggered fermions
and K\"{a}hler--Dirac fermions such a scenario is possible since the latter are invariant only under
a twisted group comprising both Lorentz and flavor symmetries~\cite{Banks:1982iq}. In staggered
approaches to QCD one
can show that the theory becomes invariant under both symmetries in the continuum limit. However this may not
be true when
taking the continuum limit in the vicinity of a strongly coupled fixed point.

Clearly, further
work, both theoretical and computational, will be required to understand these issues.
On the numerical front one will need to simulate larger systems to improve control over finite-volume effects,
determine whether there is indeed an intermediate phase, explore its nature and
measure critical exponents more precisely.  It is possible
that higher-resolution studies will reveal small but non-zero bilinear condensates on larger
volumes or that the continuous transitions we observe will become first order. Such
future studies will likely require significant improvements to the simulation algorithm,
for example by using deflation
techniques and/or carefully chosen preconditioners to handle the small fermion eigenvalues.

\acknowledgments This work is supported in part by the U.S.~Department of Energy, Office of Science, Office of High Energy Physics, under Award Number DE-SC0009998. 
We thank Venkt Ayyar, Shailesh Chandrasekharan, Poul Damgaard, Joel Giedt, Simon Hands, Jan Smit, Erich Poppitz and Bob Shrock for useful discussions at various stages of this work.
Numerical computations were performed at Fermilab using USQCD resources.

\raggedright
\bibliographystyle{apsrev}
\bibliography{su4refs}

\begin{thebibliography}{26}
\expandafter\ifx\csname natexlab\endcsname\relax\def\natexlab#1{#1}\fi
\expandafter\ifx\csname bibnamefont\endcsname\relax
  \def\bibnamefont#1{#1}\fi
\expandafter\ifx\csname bibfnamefont\endcsname\relax
  \def\bibfnamefont#1{#1}\fi
\expandafter\ifx\csname citenamefont\endcsname\relax
  \def\citenamefont#1{#1}\fi
\expandafter\ifx\csname url\endcsname\relax
  \def\url#1{\texttt{#1}}\fi
\expandafter\ifx\csname urlprefix\endcsname\relax\def\urlprefix{URL }\fi
\providecommand{\bibinfo}[2]{#2}
\providecommand{\eprint}[2][]{\url{#2}}

\bibitem[{\citenamefont{Stephenson and Thornton}(1988)}]{Stephenson:1988td}
\bibinfo{author}{\bibfnamefont{D.}~\bibnamefont{Stephenson}} \bibnamefont{and}
  \bibinfo{author}{\bibfnamefont{A.}~\bibnamefont{Thornton}},
  \bibinfo{journal}{Phys. Lett.} \textbf{\bibinfo{volume}{B212}},
  \bibinfo{pages}{479} (\bibinfo{year}{1988}).

\bibitem[{\citenamefont{Hasenfratz and Neuhaus}(1989)}]{Hasenfratz:1988vc}
\bibinfo{author}{\bibfnamefont{A.}~\bibnamefont{Hasenfratz}} \bibnamefont{and}
  \bibinfo{author}{\bibfnamefont{T.}~\bibnamefont{Neuhaus}},
  \bibinfo{journal}{Phys. Lett.} \textbf{\bibinfo{volume}{B220}},
  \bibinfo{pages}{435} (\bibinfo{year}{1989}).

\bibitem[{\citenamefont{Lee et~al.}(1990{\natexlab{a}})\citenamefont{Lee,
  Shigemitsu, and Shrock}}]{Lee:1989xq}
\bibinfo{author}{\bibfnamefont{I.-H.} \bibnamefont{Lee}},
  \bibinfo{author}{\bibfnamefont{J.}~\bibnamefont{Shigemitsu}},
  \bibnamefont{and} \bibinfo{author}{\bibfnamefont{R.~E.}
  \bibnamefont{Shrock}}, \bibinfo{journal}{Nucl. Phys.}
  \textbf{\bibinfo{volume}{B330}}, \bibinfo{pages}{225}
  (\bibinfo{year}{1990}{\natexlab{a}}).

\bibitem[{\citenamefont{Lee et~al.}(1990{\natexlab{b}})\citenamefont{Lee,
  Shigemitsu, and Shrock}}]{Lee:1989mi}
\bibinfo{author}{\bibfnamefont{I.-H.} \bibnamefont{Lee}},
  \bibinfo{author}{\bibfnamefont{J.}~\bibnamefont{Shigemitsu}},
  \bibnamefont{and} \bibinfo{author}{\bibfnamefont{R.~E.}
  \bibnamefont{Shrock}}, \bibinfo{journal}{Nucl. Phys.}
  \textbf{\bibinfo{volume}{B334}}, \bibinfo{pages}{265}
  (\bibinfo{year}{1990}{\natexlab{b}}).

\bibitem[{\citenamefont{Bock and De}(1990)}]{Bock:1990cx}
\bibinfo{author}{\bibfnamefont{W.}~\bibnamefont{Bock}} \bibnamefont{and}
  \bibinfo{author}{\bibfnamefont{A.~K.} \bibnamefont{De}},
  \bibinfo{journal}{Phys. Lett.} \textbf{\bibinfo{volume}{B245}},
  \bibinfo{pages}{207} (\bibinfo{year}{1990}).

\bibitem[{\citenamefont{Hasenfratz et~al.}(1991)\citenamefont{Hasenfratz,
  Hasenfratz, Jansen, Kuti, and Shen}}]{Hasenfratz:1991it}
\bibinfo{author}{\bibfnamefont{A.}~\bibnamefont{Hasenfratz}},
  \bibinfo{author}{\bibfnamefont{P.}~\bibnamefont{Hasenfratz}},
  \bibinfo{author}{\bibfnamefont{K.}~\bibnamefont{Jansen}},
  \bibinfo{author}{\bibfnamefont{J.}~\bibnamefont{Kuti}}, \bibnamefont{and}
  \bibinfo{author}{\bibfnamefont{Y.}~\bibnamefont{Shen}},
  \bibinfo{journal}{Nucl. Phys.} \textbf{\bibinfo{volume}{B365}},
  \bibinfo{pages}{79} (\bibinfo{year}{1991}).

\bibitem[{\citenamefont{Golterman et~al.}(1993)\citenamefont{Golterman,
  Petcher, and Rivas}}]{Golterman:1992yha}
\bibinfo{author}{\bibfnamefont{M.~F.~L.} \bibnamefont{Golterman}},
  \bibinfo{author}{\bibfnamefont{D.~N.} \bibnamefont{Petcher}},
  \bibnamefont{and} \bibinfo{author}{\bibfnamefont{E.}~\bibnamefont{Rivas}},
  \bibinfo{journal}{Nucl. Phys.} \textbf{\bibinfo{volume}{B395}},
  \bibinfo{pages}{596} (\bibinfo{year}{1993}), \eprint{hep-lat/9206010}.

\bibitem[{\citenamefont{Ayyar and Chandrasekharan}(2015)}]{Ayyar:2014eua}
\bibinfo{author}{\bibfnamefont{V.}~\bibnamefont{Ayyar}} \bibnamefont{and}
  \bibinfo{author}{\bibfnamefont{S.}~\bibnamefont{Chandrasekharan}},
  \bibinfo{journal}{Phys. Rev.} \textbf{\bibinfo{volume}{D91}},
  \bibinfo{pages}{065035} (\bibinfo{year}{2015}), \eprint{1410.6474}.

\bibitem[{\citenamefont{Ayyar and
  Chandrasekharan}(2016{\natexlab{a}})}]{Ayyar:2015lrd}
\bibinfo{author}{\bibfnamefont{V.}~\bibnamefont{Ayyar}} \bibnamefont{and}
  \bibinfo{author}{\bibfnamefont{S.}~\bibnamefont{Chandrasekharan}},
  \bibinfo{journal}{Phys. Rev.} \textbf{\bibinfo{volume}{D93}},
  \bibinfo{pages}{081701} (\bibinfo{year}{2016}{\natexlab{a}}),
  \eprint{1511.09071}.

\bibitem[{\citenamefont{Catterall}(2016)}]{Catterall:2015zua}
\bibinfo{author}{\bibfnamefont{S.}~\bibnamefont{Catterall}},
  \bibinfo{journal}{JHEP} \textbf{\bibinfo{volume}{1601}}, \bibinfo{pages}{121}
  (\bibinfo{year}{2016}), \eprint{1510.04153}.

\bibitem[{\citenamefont{He et~al.}(2016)\citenamefont{He, Wu, You, Xu, Meng,
  and Lu}}]{He:2016sbs}
\bibinfo{author}{\bibfnamefont{Y.-Y.} \bibnamefont{He}},
  \bibinfo{author}{\bibfnamefont{H.-Q.} \bibnamefont{Wu}},
  \bibinfo{author}{\bibfnamefont{Y.-Z.} \bibnamefont{You}},
  \bibinfo{author}{\bibfnamefont{C.}~\bibnamefont{Xu}},
  \bibinfo{author}{\bibfnamefont{Z.~Y.} \bibnamefont{Meng}}, \bibnamefont{and}
  \bibinfo{author}{\bibfnamefont{Z.-Y.} \bibnamefont{Lu}},
  \bibinfo{journal}{Phys. Rev.} \textbf{\bibinfo{volume}{B94}},
  \bibinfo{pages}{241111} (\bibinfo{year}{2016}), \eprint{1603.08376}.

\bibitem[{\citenamefont{Ayyar and
  Chandrasekharan}(2016{\natexlab{b}})}]{Ayyar:2016lxq}
\bibinfo{author}{\bibfnamefont{V.}~\bibnamefont{Ayyar}} \bibnamefont{and}
  \bibinfo{author}{\bibfnamefont{S.}~\bibnamefont{Chandrasekharan}},
  \bibinfo{journal}{JHEP} \textbf{\bibinfo{volume}{1610}}, \bibinfo{pages}{058}
  (\bibinfo{year}{2016}{\natexlab{b}}), \eprint{1606.06312}.

\bibitem[{\citenamefont{Fidkowski and Kitaev}(2010)}]{Fidkowski:2009dba}
\bibinfo{author}{\bibfnamefont{L.}~\bibnamefont{Fidkowski}} \bibnamefont{and}
  \bibinfo{author}{\bibfnamefont{A.}~\bibnamefont{Kitaev}},
  \bibinfo{journal}{Phys. Rev.} \textbf{\bibinfo{volume}{B81}},
  \bibinfo{pages}{134509} (\bibinfo{year}{2010}), \eprint{0904.2197}.

\bibitem[{\citenamefont{Morimoto et~al.}(2015)\citenamefont{Morimoto, Furusaki,
  and Mudry}}]{Morimoto:2015lua}
\bibinfo{author}{\bibfnamefont{T.}~\bibnamefont{Morimoto}},
  \bibinfo{author}{\bibfnamefont{A.}~\bibnamefont{Furusaki}}, \bibnamefont{and}
  \bibinfo{author}{\bibfnamefont{C.}~\bibnamefont{Mudry}},
  \bibinfo{journal}{Phys. Rev.} \textbf{\bibinfo{volume}{B92}},
  \bibinfo{pages}{125104} (\bibinfo{year}{2015}), \eprint{1505.06341}.

\bibitem[{\citenamefont{Bock et~al.}(1992)\citenamefont{Bock, Smit, and
  Vink}}]{Bock:1992yr}
\bibinfo{author}{\bibfnamefont{W.}~\bibnamefont{Bock}},
  \bibinfo{author}{\bibfnamefont{J.}~\bibnamefont{Smit}}, \bibnamefont{and}
  \bibinfo{author}{\bibfnamefont{J.~C.} \bibnamefont{Vink}},
  \bibinfo{journal}{Phys. Lett.} \textbf{\bibinfo{volume}{B291}},
  \bibinfo{pages}{297} (\bibinfo{year}{1992}), \eprint{hep-lat/9206008}.

\bibitem[{\citenamefont{van~den Doel and Smit}(1983)}]{vandenDoel:1983mf}
\bibinfo{author}{\bibfnamefont{C.}~\bibnamefont{van~den Doel}}
  \bibnamefont{and} \bibinfo{author}{\bibfnamefont{J.}~\bibnamefont{Smit}},
  \bibinfo{journal}{Nucl. Phys.} \textbf{\bibinfo{volume}{B228}},
  \bibinfo{pages}{122} (\bibinfo{year}{1983}).

\bibitem[{\citenamefont{Golterman and Smit}(1984)}]{Golterman:1984cy}
\bibinfo{author}{\bibfnamefont{M.~F.~L.} \bibnamefont{Golterman}}
  \bibnamefont{and} \bibinfo{author}{\bibfnamefont{J.}~\bibnamefont{Smit}},
  \bibinfo{journal}{Nucl. Phys.} \textbf{\bibinfo{volume}{B245}},
  \bibinfo{pages}{61} (\bibinfo{year}{1984}).

\bibitem[{\citenamefont{Eichten and Preskill}(1986)}]{Eichten:1985ft}
\bibinfo{author}{\bibfnamefont{E.}~\bibnamefont{Eichten}} \bibnamefont{and}
  \bibinfo{author}{\bibfnamefont{J.}~\bibnamefont{Preskill}},
  \bibinfo{journal}{Nucl. Phys.} \textbf{\bibinfo{volume}{B268}},
  \bibinfo{pages}{179} (\bibinfo{year}{1986}).

\bibitem[{\citenamefont{Catterall and Veernala}(2013)}]{Catterall:2013sto}
\bibinfo{author}{\bibfnamefont{S.}~\bibnamefont{Catterall}} \bibnamefont{and}
  \bibinfo{author}{\bibfnamefont{A.}~\bibnamefont{Veernala}},
  \bibinfo{journal}{Phys. Rev.} \textbf{\bibinfo{volume}{D88}},
  \bibinfo{pages}{114510} (\bibinfo{year}{2013}), \eprint{1306.5668}.

\bibitem[{\citenamefont{Witten}(1978)}]{Witten:1978qu}
\bibinfo{author}{\bibfnamefont{E.}~\bibnamefont{Witten}},
  \bibinfo{journal}{Nucl. Phys.} \textbf{\bibinfo{volume}{B145}},
  \bibinfo{pages}{110} (\bibinfo{year}{1978}).

\bibitem[{\citenamefont{Hands et~al.}(2001)\citenamefont{Hands, Kogut, and
  Strouthos}}]{Hands:2001cs}
\bibinfo{author}{\bibfnamefont{S.~J.} \bibnamefont{Hands}},
  \bibinfo{author}{\bibfnamefont{J.~B.} \bibnamefont{Kogut}}, \bibnamefont{and}
  \bibinfo{author}{\bibfnamefont{C.~G.} \bibnamefont{Strouthos}},
  \bibinfo{journal}{Phys. Lett.} \textbf{\bibinfo{volume}{B515}},
  \bibinfo{pages}{407} (\bibinfo{year}{2001}), \eprint{hep-lat/0107004}.

\bibitem[{\citenamefont{You and Xu}(2015)}]{You:2014vea}
\bibinfo{author}{\bibfnamefont{Y.-Z.} \bibnamefont{You}} \bibnamefont{and}
  \bibinfo{author}{\bibfnamefont{C.}~\bibnamefont{Xu}}, \bibinfo{journal}{Phys.
  Rev.} \textbf{\bibinfo{volume}{B91}}, \bibinfo{pages}{125147}
  (\bibinfo{year}{2015}), \eprint{1412.4784}.

\bibitem[{\citenamefont{BenTov and Zee}(2016)}]{BenTov:2015gra}
\bibinfo{author}{\bibfnamefont{Y.}~\bibnamefont{BenTov}} \bibnamefont{and}
  \bibinfo{author}{\bibfnamefont{A.}~\bibnamefont{Zee}},
  \bibinfo{journal}{Phys. Rev.} \textbf{\bibinfo{volume}{D93}},
  \bibinfo{pages}{065036} (\bibinfo{year}{2016}), \eprint{1505.04312}.

\bibitem[{\citenamefont{Wang and Wen}(2013)}]{Wang:2013yta}
\bibinfo{author}{\bibfnamefont{J.}~\bibnamefont{Wang}} \bibnamefont{and}
  \bibinfo{author}{\bibfnamefont{X.-G.} \bibnamefont{Wen}}
  (\bibinfo{year}{2013}), \eprint{1307.7480}.

\bibitem[{\citenamefont{Poppitz and Shang}(2010)}]{Poppitz:2010at}
\bibinfo{author}{\bibfnamefont{E.}~\bibnamefont{Poppitz}} \bibnamefont{and}
  \bibinfo{author}{\bibfnamefont{Y.}~\bibnamefont{Shang}},
  \bibinfo{journal}{Int. J. Mod. Phys.} \textbf{\bibinfo{volume}{A25}},
  \bibinfo{pages}{2761} (\bibinfo{year}{2010}), \eprint{1003.5896}.

\bibitem[{\citenamefont{Banks et~al.}(1982)\citenamefont{Banks, Dothan, and
  Horn}}]{Banks:1982iq}
\bibinfo{author}{\bibfnamefont{T.}~\bibnamefont{Banks}},
  \bibinfo{author}{\bibfnamefont{Y.}~\bibnamefont{Dothan}}, \bibnamefont{and}
  \bibinfo{author}{\bibfnamefont{D.}~\bibnamefont{Horn}},
  \bibinfo{journal}{Phys. Lett.} \textbf{\bibinfo{volume}{B117}},
  \bibinfo{pages}{413} (\bibinfo{year}{1982}).

\end{thebibliography}
\end{document}